\documentclass[aps,prl,letterpaper,twocolumn,superscriptaddress]{revtex4-2}
\usepackage{graphicx} 
\usepackage{float}
\usepackage{tikz}
\usepackage{amsmath}
\usepackage{amssymb}
\usepackage{physics}
\usepackage{float}

\begin{document}
\title{Long-lived revivals and real-space fragmentation in chains of multispecies Rydberg atoms}
\author{Jose Soto-Garcia}
\affiliation{  Rudolf Peierls Centre for Theoretical Physics, University of Oxford, Clarendon Laboratory, Oxford OX1 3PU, United Kingdom}
\author{Natalia Chepiga}
\affiliation{  Rudolf Peierls Centre for Theoretical Physics, University of Oxford, Clarendon Laboratory, Oxford OX1 3PU, United Kingdom}

\begin{abstract}
Arrays of Rydberg atoms provide a powerful platform for exploring constrained quantum dynamics and nonergodic many-body phenomena. While most work has focused on single-species systems, multispecies architectures offer additional interaction channels and enable new forms of dynamical constraints.
We study the nonequilibrium dynamics of one-dimensional dual-species Rydberg chains of Cs and Rb atoms with species-dependent van der Waals interactions. Using large-scale matrix product state simulations, we show that the competition between intraspecies repulsion and interspecies attraction induces dynamical fragmentation, marked by the coexistence of extended frozen regions and localized oscillatory sectors. The frozen regions act as emergent barriers that isolate and protect coherent dynamics.
In the purely repulsive regime, we find that species-selective quenches drive spontaneous fragmentation, leading to dynamically disconnected regions with irregular revivals. These phenomena are robust across interaction regimes, revealing a universal mechanism for fragmentation and establishing multispecies Rydberg arrays as a versatile platform for exploring nonequilibrium quantum dynamics beyond single-species systems.
\end{abstract}

\maketitle

\textbf{Introduction.} Arrays of neutral atoms excited to high-lying Rydberg states have emerged as a premier platform for exploring strongly correlated quantum dynamics and nonequilibrium many-body physics~\cite{keesling2019quantum, lukin2024quantum, Henriet_2020, soto2024resolving}. In these systems, atoms trapped in optical tweezer arrays can be individually controlled and arranged in programmable geometries, while excitation to Rydberg states generates strong, long-range interactions. A central feature of these architectures is the \textit{Rydberg blockade} mechanism, where the strong repulsion or attraction between nearby excitations limits simultaneous transitions to the Rydberg state within a certain distance~\cite{lukin2001dipole, browaeys2016experimental}. This blockade effectively imposes kinetic constraints on the Hilbert space, giving rise to a rich landscape of constrained quantum dynamics.

Such constraints have led to the discovery of several emergent phenomena. In the context of thermalization, one-dimensional Rydberg chains have provided the first experimental observation of quantum many-body scars---atypical non-thermal eigenstates embedded within an otherwise ergodic spectrum~\cite{turner2018weak, bernien2017probing, serbyn2021quantum, papic2022weak, zhao2025observation, datla2026statistical}. When initialized in specific product states, these scarred systems exhibit persistent, long-lived coherent revivals, defying the expectations of the eigenstate thermalization hypothesis (ETH)~\cite{ho2019periodic, daniel2023bridging}. Beyond scarring, constrained systems can exhibit Hilbert space fragmentation into exponentially many dynamically disconnected sectors due to local constraints~\cite{sala2020ergodicity, moudgalya2022thermalization, yang2025probing, kwan2025minimal}. Such fragmentation is known to stabilize localized excitations and anomalous transport, highlighting the profound role of constraints in the thermalization of quantum matter~\cite{zhao2020quantum, you2022quantum, francica2023hilbert}.

To date, much of the phenomenology of Rydberg many-body dynamics has been explored in single-species arrays, where laser driving and repulsive van der Waals (vdW) interactions realize constrained spin models such as the PXP model and its extensions~\cite{bernien2017probing,turner2018weak,fendley2004competing,lesanovsky2012interacting}. These constraints underlie quantum many-body scars and other forms of weak ergodicity breaking. More recently, facilitation~\cite{ates2007antiblockade, amthor2010evidence} has provided a complementary route beyond the blockade paradigm: by matching the laser detuning to selected interaction energies, blockade and facilitation can be combined to engineer higher-order constrained Hamiltonians exhibiting \(Z_{2k}\) scars, Hilbert-space fragmentation, Krylov-restricted thermalization, and statistically localized dynamics~\cite{zhao2025observation,datla2026statistical}.

Despite this progress, most existing realizations rely on a single atomic species. Although suitably chosen pairs of Rydberg levels within a single species can support multiple interaction channels, including resonant dipole-dipole exchange and van der Waals interactions \cite{barredo2015coherent, de2017optical, zeybek2023quantum, emperauger2025benchmarking}, independently controlling these channels and the associated dynamical degrees of freedom can be experimentally demanding, and despite individual elements of this toolbox have been demonstrated, to our knowledge the complete level of independent control required by the protocol considered here has not yet been experimentally demonstrated in a single-species Rydberg array.

Recent breakthroughs in dual-species Rydberg platforms, such as interleaved arrays of Cs and Rb atoms, provide a direct route to realizing complex constraint structures and species-dependent dynamical protocols~\cite{anand2024dual, liu2024novel, fang2025interleaved, dobrzyniecki2025tunable}. By exploiting species-dependent Rydberg levels and spectrally distinct optical transitions, the two components can be addressed independently, enabling interaction landscapes in which intraspecies repulsion coexists with attractive or selectively enhanced interspecies interactions. This expanded and highly controllable toolbox makes it possible to engineer a broad range of species-dependent interactions and constraints. Closely related functionality may also be achievable using different isotopes of the same atomic element, which can likewise serve as distinguishable atomic components and have been demonstrated in mixed-isotope tweezer arrays \cite{zeng2017entangling, sheng2022defect, wei2026enhanced}.

In this work, we investigate the nonequilibrium dynamics of dual-species Rydberg chains. First, we show that the interplay between intraspecies repulsion and interspecies attraction enables the formation of dynamically isolated clusters separated by frozen regions. We demonstrate that these regions act as effective barriers, shielding each cluster from external impurities and protecting its coherent time evolution. Finally, we show that species-selective quenches induce spontaneous real-space fragmentation of Rydberg arrays even in the purely repulsive regime, providing a novel route to control and manipulate quantum processors.

\textbf{The model.} We consider one-dimensional arrays composed of two atomic species, Cs and Rb, following the dual-species Rydberg platform architecture outlined in Ref.~\cite{anand2024dual}. Each atom is modeled as an effective two-level system consisting of a ground state $\ket{g}$ and a highly excited Rydberg state $\ket{r}$. The many-body dynamics of the system is governed by the Hamiltonian

\begin{equation}
H = \sum_{\alpha=\mathrm{Cs},\mathrm{Rb}} \sum_{i\in\alpha} \left( - \frac{\Omega_{\alpha}}{2}\sigma_i^{x} - \Delta_{\alpha} n_i \right) 
+ \sum_{i<j} \frac{C_{ij}}{|i-j|^{6}} n_i n_j.
\end{equation}
The operator $n_i = \ket{r}_i\bra{r}_i$ measures the Rydberg excitation on site $i$, while $\sigma_i^x = \ket{g}_i\bra{r}_i + \ket{r}_i\bra{g}_i$ drives coherent transitions between the ground and Rydberg states. The parameters $\Omega_\alpha$ and $\Delta_\alpha$ denote the Rabi frequency and detuning for a species $\alpha\in\{\mathrm {Rb, Cs}\}$. Throughout this work, we set $\Omega_{\mathrm{Cs}} = \Omega_{\mathrm{Rb}} = 1$ and $\hbar = 1$.
The final term accounts for the vdW interactions between Rydberg atoms, where the coefficient $C_{ij}$ is determined by the species occupying sites $i$ and $j$. For atoms of the same species, we consider repulsive interactions $C_{\alpha\alpha} \equiv C_0 > 0$. For interspecies pairs (Cs--Rb), the interaction coefficient $C_{\alpha\beta} \equiv C_1 $ is tuned to be either repulsive or attractive, and we discuss both regimes below. In the main text we focus on vdW interspecies coupling, while we generalize these conclusions to F\"orster resonance in the End Matter and to other combinations of interactions in the Supplementary Material.

We study the quench dynamics of this model numerically using large-scale tensor-network simulations based on matrix product states (MPS)~\cite{schollwock2011density}. We prepare initial states with density matrix renormalization group algorithm (DMRG)~\cite{white1992density,schollwock2011density}. We then quench the system by switching off the corresponding laser detuning(s)  $\Delta_\alpha$ and model this with time-dependent variational principle (TDVP)~\cite{haegeman2011time, haegeman2016unifying, paeckel2019time}. Additional technical details of the algorithms are provided in the End Matter.

\begin{figure}[h]
    \centering
    \includegraphics[width=\linewidth]{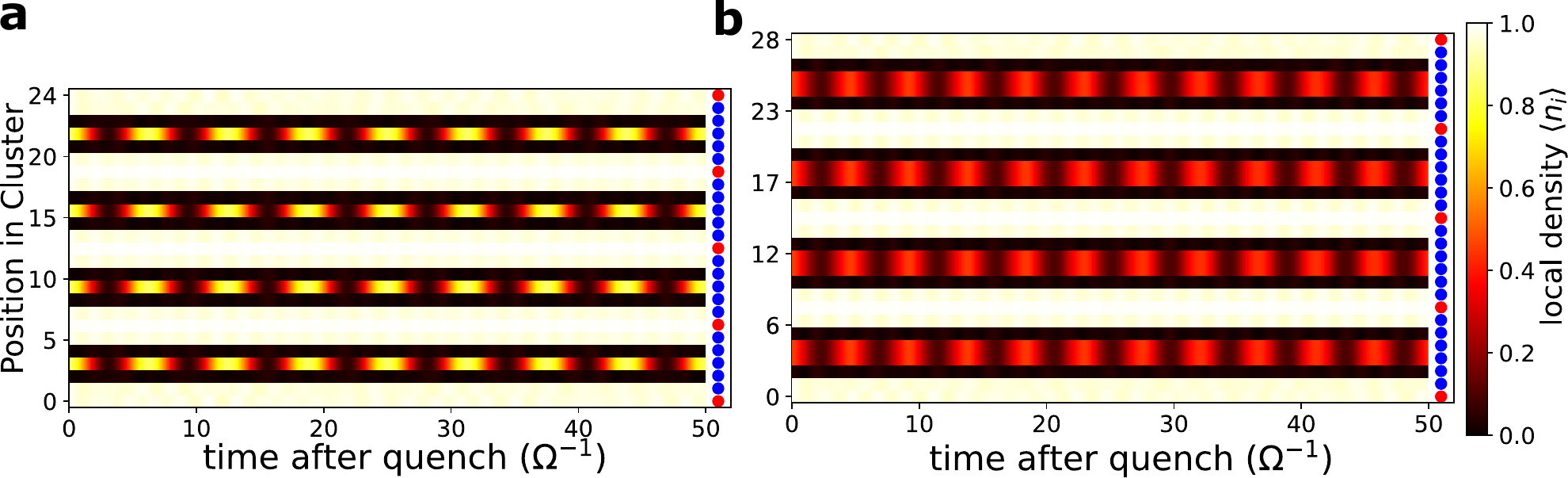}
    \caption{Time evolution of a local density after a quantum quench from a crystalline initial state with attractive interspecies and repulsive intraspecies vdW interactions in arrays with  (a) $\mathrm{Cs}$–$\mathrm{Rb}_5$ and (b)  $\mathrm{Cs}$–$\mathrm{Rb}_6$. Attractive Cs--Rb interactions pin Rb--Cs--Rb clusters, producing extended frozen regions and reducing dynamically active fragments to (a)  a single active Rb atom undergoing coherent oscillations and (b) two neighboring Rb atoms forming an oscillating resonant pair.  
    Sidebars: Red (blue) circles indicate Cs (Rb) atoms. Numerically extracted periods $T$ of oscillations coincide with theoretical predictions~\cite{bernien2017probing} for single and pairs of atoms within 1.4\%.}
    \label{fig:singleoscillation}
\end{figure}

\textbf{Chain fragmentation via attractive interspecies interactions.} We first focus on the simplest array with built-in periodic pattern in which a single Cs atom is followed by $n$ Rb atoms~\footnote{We add another Cs atom at the end of the chain to realize symmetric boundary conditions in an open chain}; for simplicity we will refer to such a configuration as  $\mathrm{Cs}$–$\mathrm{Rb}_n$.
The interplay between intraspecies repulsion and interspecies attraction generates effective kinetic constraints that strongly restrict the accessible Hilbert space. The resulting nonequilibrium dynamics features real-space fragmentation as shown in Fig.~\ref{fig:singleoscillation}. We initiate the evolution via a global quantum quench, starting from a crystalline ground state prepared at $\Delta_{\textnormal{Cs}} = \Delta_{\textnormal{Rb}} = 1.2\,\Omega$ with interaction strengths set to $C_{0} = -C_{1} = 1.2^6\,\Omega$. At $t=0$, the detunings are abruptly switched off ($\Delta_{\textnormal{Cs}} = \Delta_{\textnormal{Rb}} = 0$), triggering coherent evolution of the interacting Hamiltonian. 

Let us first consider the dual-species pattern $\mathrm{Cs}$–$\mathrm{Rb}_5$, where attractive Cs–Rb interactions bind Cs excitations to their nearest Rb neighbors, forming extended high-density frozen regions that appear as light stripes in Fig.~\ref{fig:singleoscillation}(a). The dynamics is therefore confined to the remaining active sectors, which, for $\mathrm{Cs}$–$\mathrm{Rb}_5$, consist of a single Rb atom undergoing coherent Rabi oscillations. This active atom is bounded on both sides by atoms that remain in the ground state throughout the quench, corresponding to the black regions in Fig.~\ref{fig:singleoscillation}(a). For the larger unit cell $\mathrm{Cs}$–$\mathrm{Rb}_6$, the high-density Rb--Cs--Rb walls instead isolate a pair of neighboring Rb atoms that share a resonating single Rydberg excitation, again bounded by atoms that remain in the ground state, as shown in Fig.~\ref{fig:singleoscillation}(b). We extract the oscillation period $T$ by averaging the temporal separation between successive peaks over the full time window shown in the figure. The error is calculated as $1.96 \times$ standard error.  For a single active Rb atom, we obtain $T_{\mathrm{single}} \approx (6.37 \pm 0.05)\Omega^{-1}$, in agreement with the single-atom prediction $T=2\pi/\Omega$ within a relative error of $1.4\%$. When two neighboring Rb atoms form the active region, the oscillation period decreases to $T_{\mathrm{pair}} \approx (4.42 \pm 0.15)\Omega^{-1}$. The relation $T_{\mathrm{pair}} \approx T_{\mathrm{single}}/\sqrt{2}$ indicates a $\sqrt{2}$ enhancement of the effective Rabi frequency, consistent with the collective dynamics expected for two coupled resonant atoms, with a relative deviation of only $0.5\%$~\cite{bernien2017probing}.

We attribute the remaining small deviations from the ideal oscillation periods to the influence of the surrounding atoms on the local dynamics. As discussed in the Supplementary Material, neither the single-atom nor the two-atom oscillations exhibits complete population transfer between the ground and Rydberg states: the excitation density does not span the ideal ranges $0$--$1$ for the single active atom and $0$--$1/2$ per atom for the active pair. This incomplete population transfer reflects residual coupling to the surrounding frozen regions and provides a plausible explanation for the small discrepancies between the measured periods and the idealized predictions.

\begin{figure}[h]
\centering
\includegraphics[width=\linewidth]{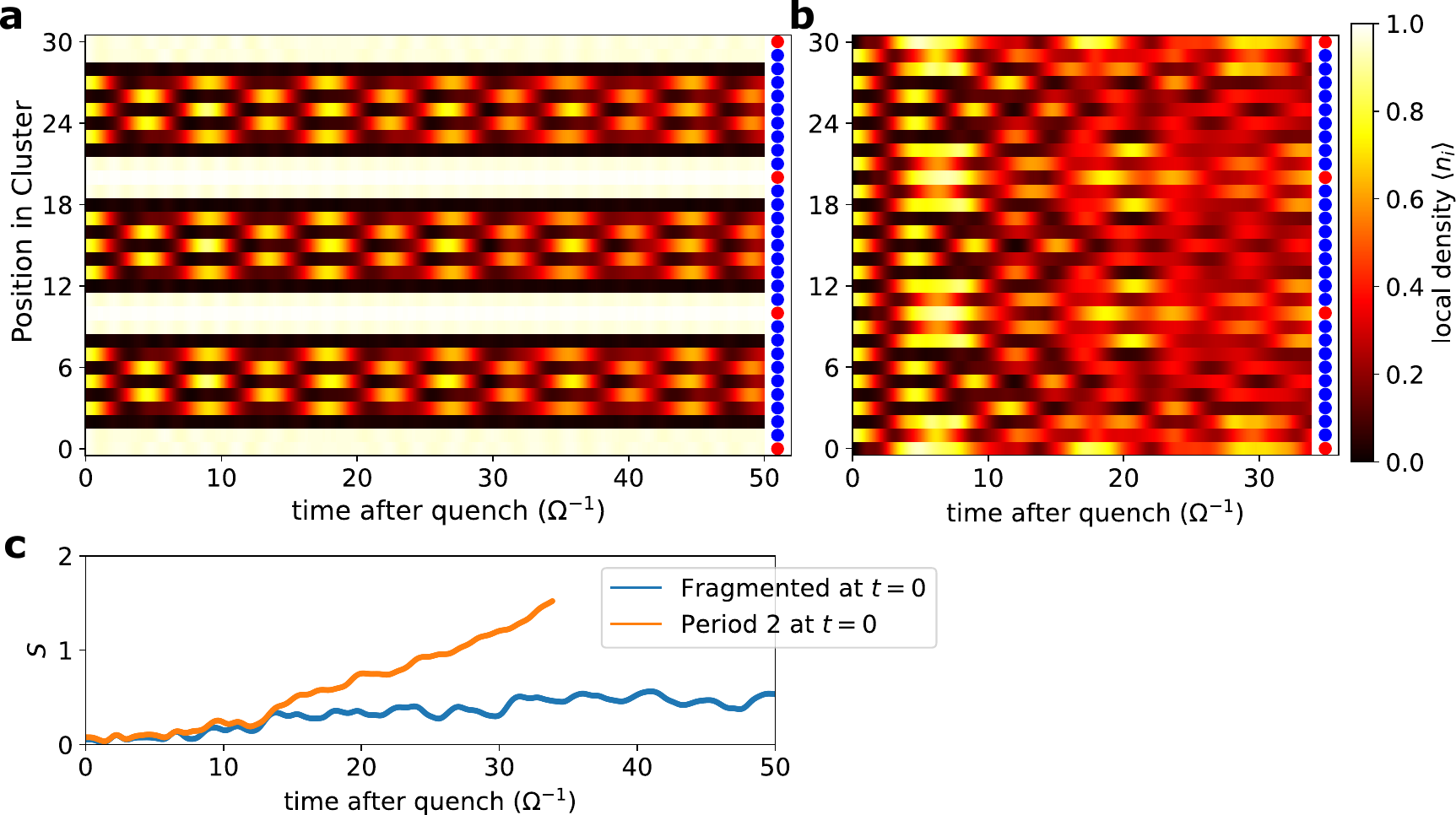}
\caption{Time evolution of Cs--Rb$_9$ array after a quantum quench   from two different initial states. (a) Coherent dynamics of local density of three few-body fragments lined between frozen high-density regions. (b) Initialized in the period-two state, the Rb--Cs--Rb cluster acts as an impurity: excitation of the Cs atom pins its neighbors, creating a defect that propagates along the chain and progressively suppresses coherent oscillations.
(c) Entanglement entropy of the systems shown in (a) (blue) and (b) (orange) bipartite across the middle of the chain. The propagating impurity induces significantly faster entanglement growth.}
\label{fig:fragmentedp2}
\end{figure}

Further varying the number of Rb atoms between Cs sites, we realize dynamically active few-body systems separated by frozen segments, as shown in Fig.~\ref{fig:fragmentedp2}(a). In the example shown, three clusters of seven Rb atoms each evolve independently, resulting in stable coherent dynamics over extended times. This is also reflected in the bipartite entanglement entropy (across the center of the chain), computed from the Schmidt values $\lambda_i$ of the matrix product state as $S = -\sum_i \lambda_i^2 \log \lambda_i^2$,
which shows no signature of divergence in Fig.~\ref{fig:fragmentedp2}(c). In the End Matter, we also present examples of long-lived revivals in extended clusters with an number of even sites, featuring oscillating resonant densities, as well as multiplex dynamics observed in arrays with unevenly distributed Cs impurities.

Until now, we have considered the cases in which the systems were initiated in a ground state of the interacting system with high-density clusters near Cs impurity appearing at $t=0$. However, a qualitatively different dynamics emerges when the system is initialized in a period-two ordered state, as shown in Fig.~\ref{fig:fragmentedp2}(b). This state is prepared as the ground state at $\Delta_{\textnormal{Cs}}=-20\,\Omega$, $\Delta_{\textnormal{Rb}}=1.2\,\Omega$, and $C_0 = -C_1 = 1.2^6\,\Omega$~\footnote{In experiments this can be done with a light-shift technique}, effectively embedding the Rb--Cs--Rb configuration as a mobile impurity within the period-two order. During the post-quench evolution presented in Fig.~\ref{fig:fragmentedp2}(b), the excitation of the central Cs atom periodically pins its Rb neighbors, locally disrupting the background's oscillatory dynamics. The resulting defect propagates ballistically through the lattice, leading to a rapid growth of entanglement entropy (see Fig.~\ref{fig:fragmentedp2}(c)) and facilitating thermalization.

\begin{figure}[h]
    \centering
    \includegraphics[width=\linewidth]{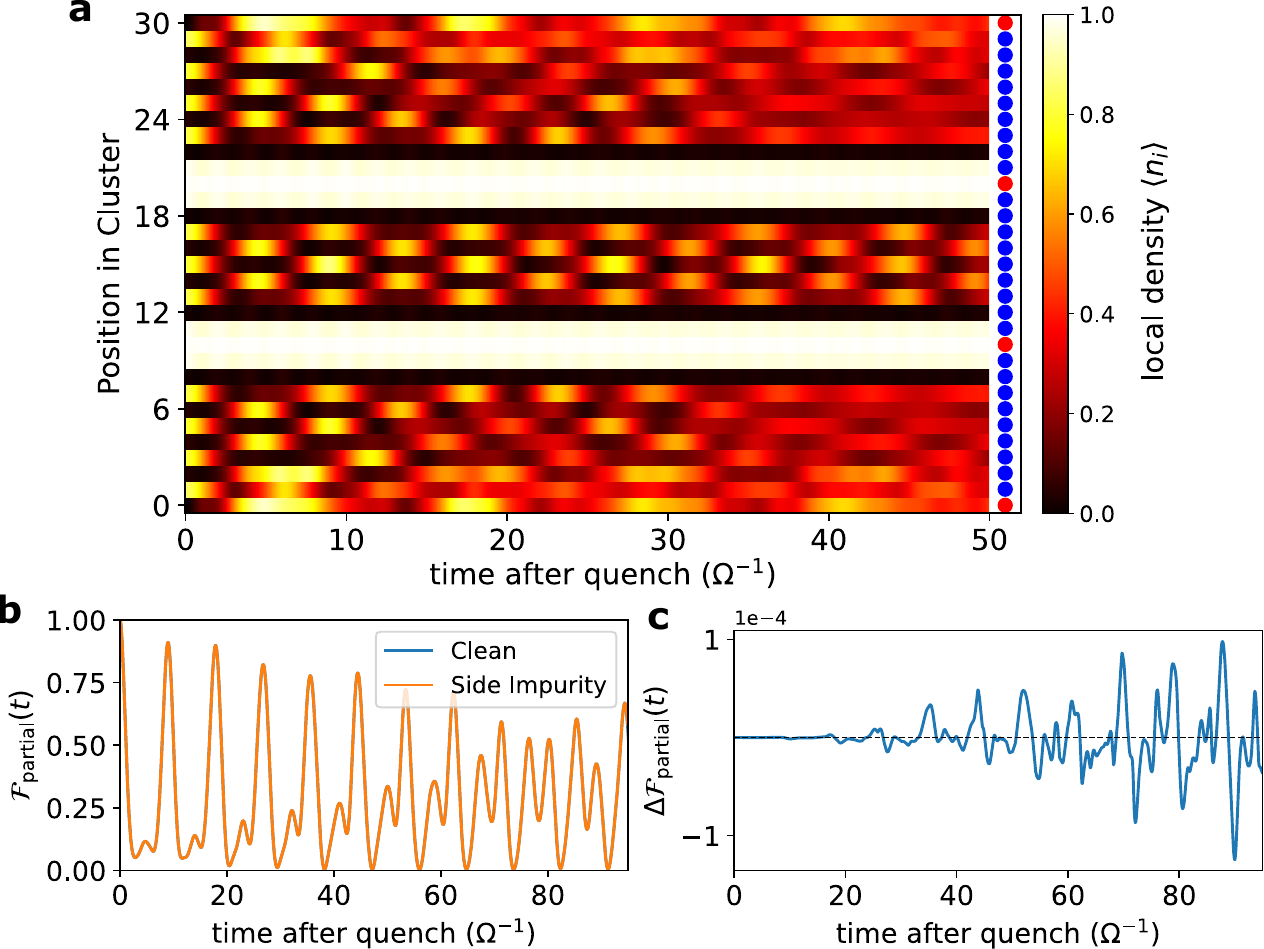}
    \caption{Robustness of dynamically active regions against external perturbations. (a) Time evolution of the Rydberg excitation density in the presence of impurities (defects) localized at the edges of the chain at $t=0$. (b) Comparison of the fidelity on the central fragment (atoms 13 to 17) with respect to the initial state for the clean system described in Fig.~\ref{fig:fragmentedp2}(a) (blue) and system with impurities at the side (orange). (c) Difference for the fidelities calculated in (b) between both systems. Fully overlapping trajectories demonstrate that the central fragment remains dynamically isolated; the propagating defects are effectively screened by the frozen regions.
    }
    \label{fig:impurity_effect}
\end{figure}

{\bf Impurity shielding.}
The dynamically active regions separated by frozen few-site walls, identified in Fig.~\ref{fig:singleoscillation} and Fig.~\ref{fig:fragmentedp2}(a), suggest a fragmentation of the array into effectively independent subsystems. A key question for experimental applications is whether these fragments are truly isolated or remain susceptible to perturbations from their surroundings. To probe this robustness, we introduce external impurities in the form of propagating defects outside a central active fragment. The Cs atoms at the edges are in the ground state, following the period-two structure of their surroundings, instead of the pinned Rydberg-Rydberg structure of Fig.~\ref{fig:fragmentedp2}(a). In practice, this is implemented by preparing an initial state with strongly shifted laser detunings near the edges of the array, resulting in the density profiles shown in Fig.~\ref{fig:impurity_effect}(a). The thermalizing dynamics induced by the boundary defects remain confined to the outer fragments, leaving the dynamics of the central segment essentially unaffected.

This shielding effect is confirmed numerically by comparing the fidelity with respect to the initial state within the central fragment, $\mathcal{F}_{\mathrm{partial}}(t)$, as shown in Fig.~\ref{fig:impurity_effect}(b). The difference between the fidelity curves with and without boundary impurities remains below $10^{-4}$ throughout the entire simulation time, as shown in Fig.~\ref{fig:impurity_effect}(c). In contrast to a generic thermalizing system, in which excitations spread and scramble local information, the multispecies constraints generate an effective barrier that strongly suppresses information transport. These results indicate that the dynamically active regions are not merely approximately isolated, but instead behave as effectively protected dynamical sectors of the Hilbert space. Interestingly, $\mathcal{F}_{\mathrm{partial}}(t)$ exhibits a two-scale oscillatory structure: fast oscillations, which appear to be associated with quantum many-body scar dynamics, together with a slower modulation of the oscillation envelope. We attribute this beating-like behavior to the presence of two nearby oscillation frequencies arising from imperfect revivals. A more detailed analysis is provided in the Supplementary Material.

\textbf{Emergent fragmentation with species-selective quenches.}
A key advantage of the multispecies architecture is the ability to address and manipulate each atomic species independently, in contrast to single-species platforms with multiple internal components. This additional degree of control enables \textit{species-selective quenches}, in which the dynamical constraints acting on one species are modified while the other remains off-resonant or pinned.

\begin{figure}[h]
    \centering
    \includegraphics[width=\linewidth]{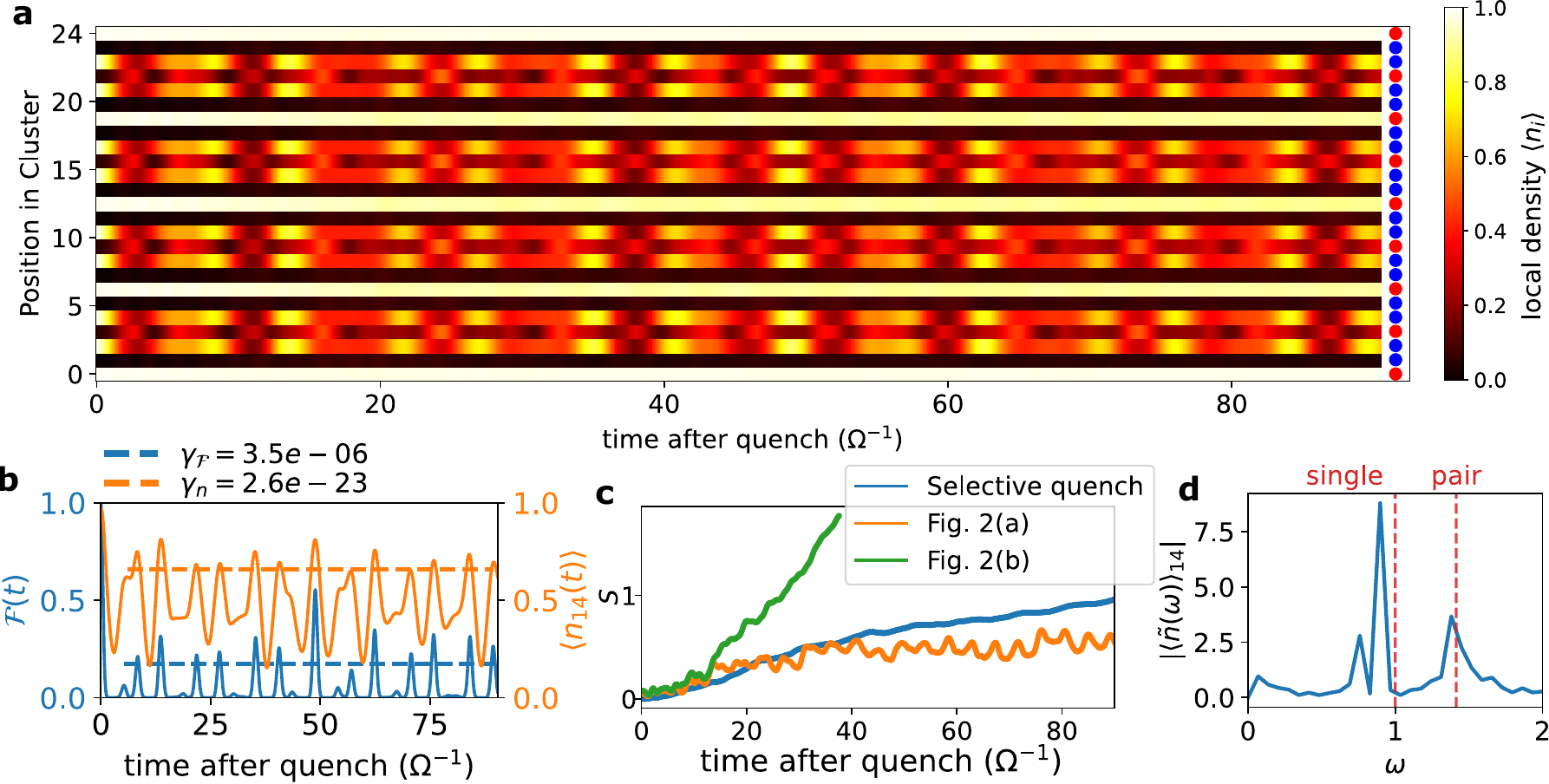}
    \caption{(a) Time evolution of the Rydberg excitation density following a selective quench of the Rb atoms from an initial period-two ordered state. The dynamics generates a spatial structure in which alternating Cs atoms remain pinned in the Rydberg state, while neighboring Rb atoms stay predominantly in the ground state, forming frozen regions that separate dynamically active segments. Within these active regions, the excitation density exhibits irregular revivals, reflecting constrained many-body dynamics. (b) Quantum fidelity $\mathcal{F}(t)$ (blue, left axis) and the local Rydberg excitation density $\langle n_{14}(t)\rangle$ at site 14 (orange, right axis). Both quantities exhibit irregular revivals rather than a single periodic oscillation. The dashed line shows an exponential fit to the local maxima, with $\gamma$ denoting the decay rate. (c) Entanglement entropy after the selective quench (blue), compared with the dynamics for the systems shown in Fig.~\ref{fig:fragmentedp2}(a) (orange) and Fig.~\ref{fig:fragmentedp2}(b) (green). The selective quench leads to substantially slower entanglement growth than the latter case and similar to the quench protocol of Fig.~\ref{fig:fragmentedp2}(a). (d) Fourier spectrum of the local excitation dynamics shown in (b). The spectrum contains several frequency components, with prominent peaks lying close to the characteristic oscillation frequencies of an isolated atom (``single'') and a resonant two-atom pair (``pair''), indicated by the red dashed lines.
}
    \label{fig:fragmentation}
\end{figure}

We apply the selective quench to an array of $\mathrm{Cs}$--$\mathrm{Rb}_2$ atoms now with all-repulsive interactions $C_0 = C_1 = 1.2^6\,\Omega$ initially prepared  in a period-two ordered state  (as a ground state at $\Delta_{\textnormal{Cs}} = \Delta_{\textnormal{Rb}} = 4\,\Omega$). By quenching only the detuning of the Rb atoms we observe a {\it spontaneous fragmentation} of the array into clusters of Rb--Rb--Cs--Rb--Rb atoms separated by (almost) frozen Cs sites as demonstrated in Fig.~\ref{fig:fragmentation}(a). 

Within each fragment, the local Rydberg excitation density \(n_i(t)\) and the quantum fidelity $\mathcal{F}(t)$ exhibit irregular, non-periodic oscillations, as shown in Fig.~\ref{fig:fragmentation}(b). The oscillations do not show any sign of damping for the duration of the simulation, as shown by the small decay rate obtained by fitting to $Ae^{-\gamma t}$, indicated as dashed line in the figure. Furthermore, as shown in  Fig.~\ref{fig:fragmentation}(c), the entanglement entropy grows substantially more slowly than in the thermalizing case shown in Fig.~\ref{fig:fragmentedp2}(b). Its growth rate is instead comparable to that observed for the system exhibiting periodic revivals in Fig.~\ref{fig:fragmentedp2}(a). The absence of a single dominant oscillation frequency in the Fourier transform of the density oscillations depicted in Fig.~\ref{fig:fragmentation}(d) indicates that the selective quench generates dynamics involving multiple characteristic frequencies within the fragmented sectors, leading to complex interference patterns rather than simple few-body periodic motion.

\section{Discussion}

To summarize, we demonstrate that multispecies Rydberg platforms significantly extend the landscape of constrained quantum dynamics by providing additional experimentally accessible control parameters beyond conventional single-Rydberg-level implementations.
For attractive interspecies interactions we report a remarkable self-organization of the dynamics into coexisting regions with sharply distinct behavior: extended frozen domains, and small active clusters undergoing coherent oscillations. As demonstrated in our results, the size and structure of these active regions are precisely controlled by the spatial arrangement of the atomic species.

The robustness of these phenomena to residual long-range interactions and external perturbations is crucial for their experimental realization. Our results show that the dynamically isolated sectors remain remarkably stable even in the presence of defects in neighboring regions of the chain. In particular, the central active fragments are effectively shielded from external impurities, with their local coherent dynamics remaining essentially unchanged over the simulated timescales. This dynamical isolation also enables the simultaneous realization of processes occurring on otherwise incompatible timescales in programmable simulators, as demonstrated in Fig.~\ref{fig:otherpatterns} of the End Matter. Such strong decoupling suggests that multispecies Rydberg arrays provide an already experimentally accessible and particularly promising platform for observing robust revival dynamics in {\it interacting} quantum systems, allowing local nonequilibrium signatures to persist well within the coherence times accessible in current experiments~\cite{bernien2017probing, keesling2019quantum, browaeys2020many, ebadi2021quantum, anand2024dual}.

Importantly, we have shown that structured constrained dynamics can also be engineered through selective quenches of individual atomic species. The ability to address Cs and Rb atoms independently enables the creation of highly nontrivial spatiotemporal patterns. Depending on the driving protocol, the dynamics can range from irregular revival behavior within isolated clusters to correlated propagation patterns characterized by ballistic excitation fronts, as we show in the End Matter. These results highlight the versatility of multispecies platforms in designing nonequilibrium dynamics with both localized and propagating features.

Throughout the main text, we consider van der Waals (vdW) potentials for both inter- and intraspecies interactions, as the off-resonant vdW regime is generally less sensitive to residual electric fields than interactions tuned close to a F"orster resonance~\cite{anand2024dual}. However, in many experimental settings, F"orster resonances provide the dominant interspecies coupling mechanism~\cite{ravets2014coherent, anand2024dual, dobrzyniecki2025tunable, white2026quantum}. Importantly, the phenomena reported here persist when the interspecies interaction is mediated by a F"orster resonance, as demonstrated by the supporting data presented in the End Matter. Furthermore, while the main text focuses on a representative set of interaction strengths for $\mathrm{Cs}$--$\mathrm{Cs}$, Cs--Rb, and Rb--Rb pairs, in the Supplementary Material we show that other choices of these couplings yield the same qualitative dynamical behavior, provided that they preserve the same hierarchy of constraints. Taken together, these observations demonstrate that the phenomena discussed here do not rely on a fine-tuned interaction potential or a specific set of interaction strengths. Rather, they are governed primarily by the interplay among the initial state, the spatial arrangement of the two species, and the quench protocol.

In conclusion, we demonstrate that multispecies Rydberg arrays provide a versatile platform for engineering constrained nonequilibrium dynamics in a highly controllable setting. The interplay of species-dependent interactions and selective control enables the emergence—either spontaneous or engineered—of dynamically isolated regions with distinct coherent behavior, offering a concrete route to dynamical fragmentation. These mechanisms are directly realizable in existing dual-species Rydberg platforms.

Recent experiments on two-dimensional heterogeneous Rydberg arrays~\cite{PhysRevX.12.011040} naturally motivate extensions of our work beyond one dimension. In particular, these systems enable the study of collective dynamics in coupled chains and ladders with barriers separating them, providing a pathway to interpolate between one- and two-dimensional out-of-equilibrium phenomena, for instance, by generating barriers separating single arrays.

\begin{acknowledgments}
We acknowledge useful discussions with J. Pritchard and S. A. Parameswaran. This work was funded by the European Union through the ERC grant (TRANGINEER, 101220181). The views and opinions expressed are those of the authors only and do not necessarily reflect those of the European Union or the European Research Council Executive Agency; neither the European Union nor the granting authority can be held responsible for them. N. C. acknowledges support from the Royal Society (grant number URFR1251326).
\end{acknowledgments}
\bibliography{bibliography}

\clearpage

\section{End Matter}

\textbf{Technical details of the algorithms.} Numerical simulations were performed using state-of-the-art density matrix renormalization-group (DMRG) methods in the matrix product state (MPS) framework~\cite{white1992density,schollwock2011density}. The algebraically decaying interactions were approximated by a matrix product operator (MPO) using a modified incremental singular value decomposition (ISVD) scheme~\cite{lin2021isvd}. We used a truncation rank $M = 19$, corresponding to an MPO bond dimension $D = M + 2 = 21$, and an ISVD tolerance of $10^{-9}$. With these parameters, the relative approximation error remained below $10^{-11}$, quantified by the cost function
$J_F = \frac{\sum_{i<j}\left(V_{ij}-\tilde V_{ij}\right)^2}{\sum_{i<j}V_{ij}^2}$,
where $V_{ij}$ denotes the exact interactions and $\tilde V_{ij}$ their MPO approximation.

Ground states were obtained using two-site MPS-based DMRG~\cite{white1992density,schollwock2011density}. Singular values were truncated at $\chi \geq 10^{-7}$, with a maximum bond dimension $D = 400$. Convergence was assumed when the relative energy difference between successive sweeps fell below $10^{-9}$.

The nonequilibrium dynamics were simulated using the two-site time-dependent variational principle (TDVP)~\cite{haegeman2011time, haegeman2016unifying, paeckel2019time} formulated in MPS. For each discrete time step $\delta t$, the local Hilbert space was updated using a Lanczos eigensolver, with a convergence criterion of $10^{-8}$ for the energy difference between successive iterations. Throughout our simulations, we employed a fixed time step of $\delta t = 0.01\,\Omega^{-1}$.

\begin{figure}[b!]
    \centering
    \includegraphics[width=0.8\linewidth]{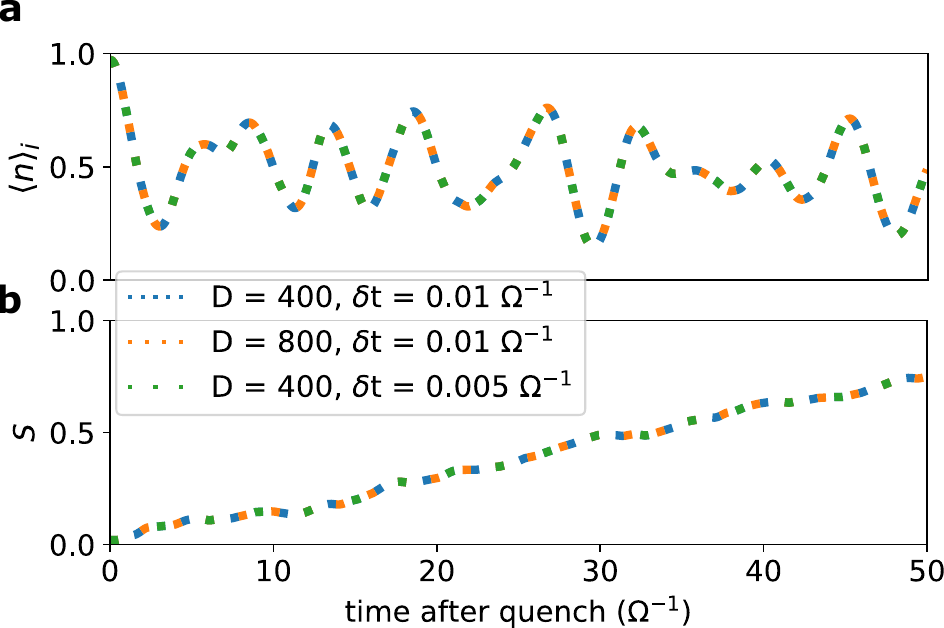}
    \caption{Numerical convergence and stability of the TDVP simulations. (a) Time evolution of the local density $n_i(t)$ at site $i=10$ and (b) growth of the bipartite entanglement entropy $S$ following the selective quench protocol described in Fig.~\ref{fig:fragmentation} of the main text. The results are compared across different bond dimensions and integration time steps listed in the legend.}
    \label{fig:bond_dimension}
\end{figure}
    The maximum bond dimension was set to $D = 400$ with a truncation threshold of $\chi = 10^{-7}$. We performed rigorous convergence checks on $D$ and $\delta t$ as demonstrated in Fig.~\ref{fig:bond_dimension}, where we compare the time evolution of the Rydberg excitation density and entanglement entropy upon increasing the bond dimension or decreasing the time step. The near-perfect overlap of the observables across chosen parameters confirms that $D=400$ and $\delta t=0.01\,\Omega^{-1}$ provide sufficient accuracy.
    
\begin{figure}[h!]
    \centering
    \includegraphics[width=0.9\linewidth]{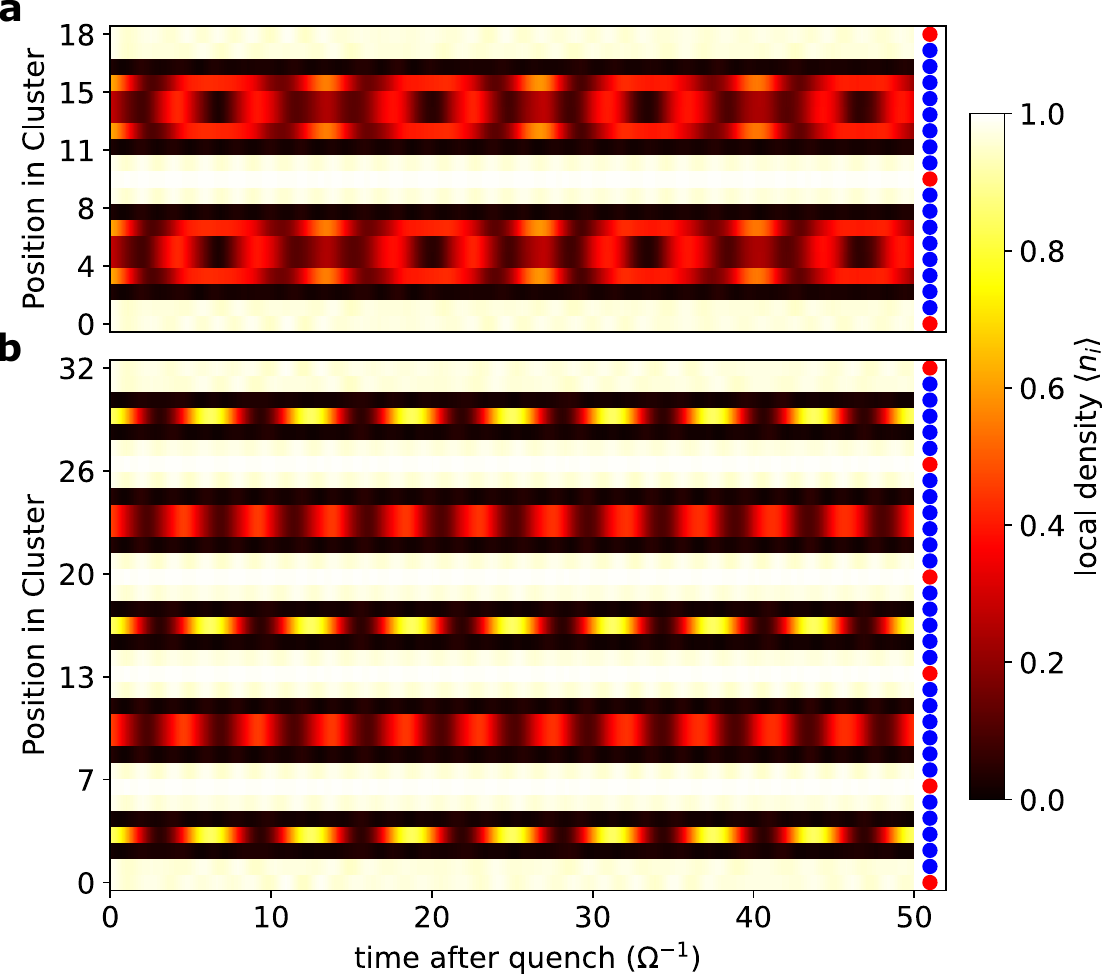}
    \caption{Diverse constrained phases and spatial multiplexing via unit-cell engineering. (a) Time evolution of the local density $n_i(t)$ for a $\mathrm{Cs}$–$\mathrm{Rb}_8$ array. A bimodal revival pattern with alternating high- and low-amplitude time intervals over a fragment of six active sites should be contrasted with Figs.~\ref {fig:singleoscillation} and \ref{fig:fragmentedp2} of the main text.
 (b) Spatiotemporal evolution of $n_i(t)$ for a heterogeneous array with alternating $\mathrm{Cs}$–$\mathrm{Rb}_5$ and $\mathrm{Cs}$–$\mathrm{Rb}_6$ unit cells. This configuration exhibits spatial multiplexing, with single-atom oscillations and resonant-pair dynamics coexisting in dynamically isolated sectors. Sidebars show dual-species arrangements with red (blue) circles denoting Cs (Rb) atoms.}
    \label{fig:otherpatterns}
\end{figure}

\textbf{Diverse Constrained Phases and Dynamical Multiplexing.} 
Here we present more complex constrained phases complementing the examples discussed in the main text. By taking the dual-species pattern $\mathrm{Cs}$--$\mathrm{Rb}_8$, we observe in Fig.~\ref{fig:otherpatterns}(a) the emergence of more complex bimodal revival patterns within the inter-pinned regions. We interpret these complex revivals as an interplay of strong revivals that we observed previously in the main text and a mismatch between the preferred ordering pattern and the finite length of the active fragment.

Furthermore, multispecies architectures with built-in fragmentation enable spatially heterogeneous arrays. In Fig.~\ref{fig:otherpatterns}(b), we demonstrate a \textit{multiplexed} configuration with alternating $\mathrm{Cs}$–$\mathrm{Rb}_5$ and $\mathrm{Cs}$–$\mathrm{Rb}_6$ unit cells along the lattice. In this regime, single-atom and resonant-pair oscillations coexist in distinct, dynamically isolated sectors. The ability to alternate unit-cell geometries within a single quench protocol highlights the potential of dual-species Rydberg chains for parallelized quantum simulations.

\begin{figure}[t]
    \centering
    \includegraphics[width=\linewidth]{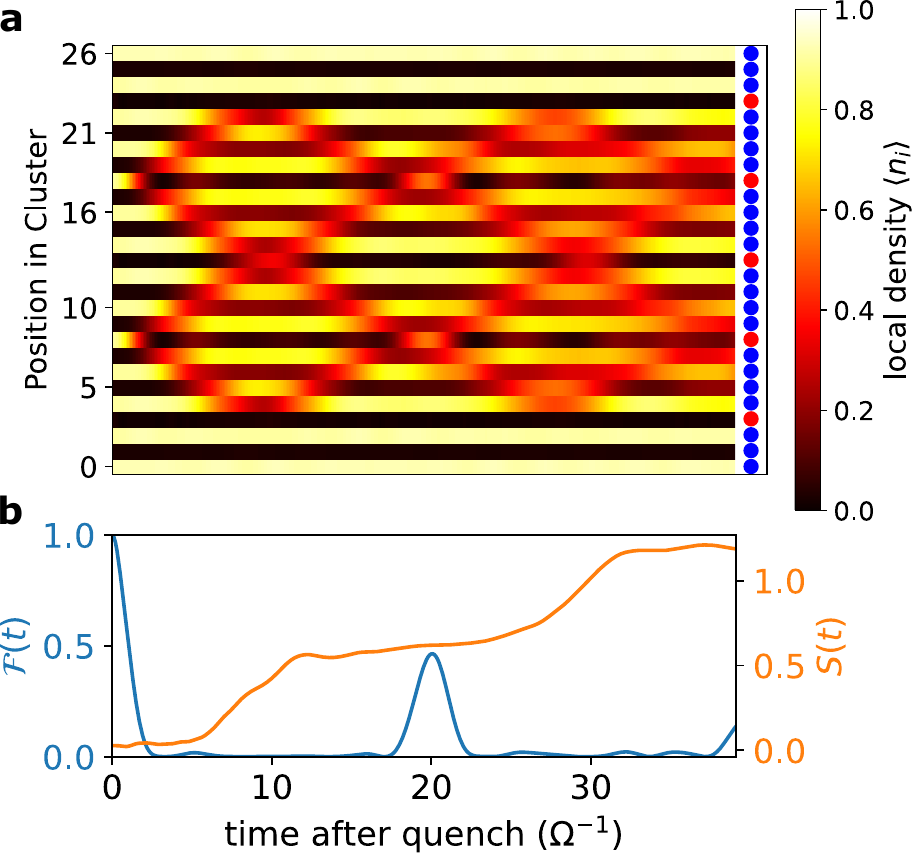}
    \caption{Emergent propagation dynamics in a $\mathrm{Cs}$--$\mathrm{Rb}_4$ array. (a) Local density following a selective quench of the Cs species features ballistic propagation and subsequent rebound of excitation fronts, resulting in a structured spatiotemporal pattern. (b) Fidelity $\mathcal{F}(t)$ with respect to the initial state (blue, left axis) and bipartite entanglement entropy $S$ (orange, right axis). The entanglement entropy shows a step-like growth, with pronounced increases coinciding with excitation waves in the middle of the chain, indicating the buildup of quantum correlations during these events.}
    \label{fig:squared}
\end{figure}

\textbf{Ballistic propagation and constrained dynamics.}
Beyond localized fragmentation, selective quench protocols enable the emergence of structured dynamical patterns governed by the underlying constraints. We consider a $\mathrm{Cs}$–$\mathrm{Rb}_4$ array, where only the Cs atoms are quenched while the Rb atoms remain in the ground state. As shown in Fig.~\ref{fig:squared}(a), this selective drive generates domain-wall-like excitations that propagate ballistically through the Rb blocks. Upon collision with excitations from neighboring unit cells, they do not thermalize but instead undergo a characteristic rebound, producing a structured, periodic spatiotemporal pattern. This behavior reflects species-dependent constraints that effectively discretize the dynamics.

The global dynamics, shown in Fig.~\ref{fig:squared}(b), exhibit fidelity revivals, indicating recurrent proximity to the initial state. The entanglement entropy displays a step-like growth, with pronounced increases synchronized with the passage of excitation fronts through the observation site. This correlation highlights the role of propagating excitations in generating quantum correlations.

\textbf{Interspecies interactions: F\"orster resonance.} In the main text, we assumed a vdW ($r^{-6}$) scaling for both intra- and interspecies interactions. In this section, we demonstrate that the emergent constrained dynamics and fragmentation properties  are robust against variations in the specific form of the interaction potential. Specifically, we consider an experimentally inspired setting where interspecies coupling is mediated by F\"orster resonance, which yields a dipolar interaction scaling as $r^{-3}$ at moderate distances~\cite{ravets2014coherent, browaeys2016experimental}.

F\"orster resonance typically involves the coupling of two different Rydberg states, resulting in a potential with both attractive and repulsive branches. Experimentally, a specific interaction branch can be targeted through either narrow-band spectroscopic selection or adiabatic state preparation via Stark-tuned electric field sweeps, allowing for the deterministic realization of either attractive or repulsive interspecies potentials~\cite{ravets2014coherent, beterov2015rydberg, weber2017calculation}. Here, we simulate the nonequilibrium dynamics using an interspecies potential $V_{\text{inter}}(r) = C_2 / r^3$, while maintaining the repulsive $r^{-6}$ intraspecies vdW interaction. For a direct comparison, we set $|C_2|= C_0$ so that the nearest-neighbor coupling remains consistent with the values used in the main text.

\begin{figure*}[b]
    \includegraphics[width=\textwidth]{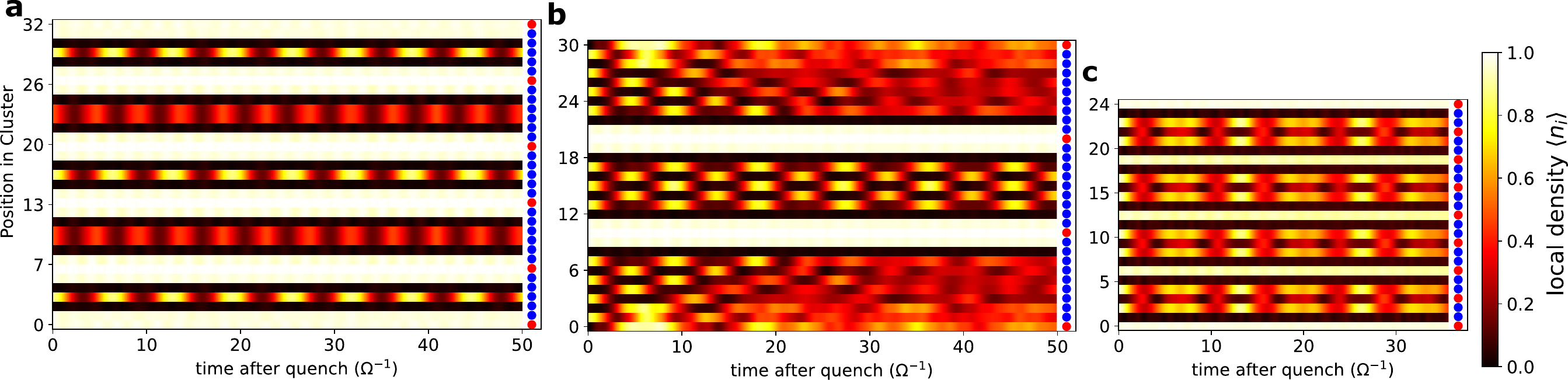}
    \caption{Time evolution of local density for the three primary protocols discussed in this paper
    for arrays with interspecies F\"orster coupling decaying with distance as $1/r^3$. The results in (a), (b) and (c) should be compared with the results for vdW interactions presented in Fig.~\ref{fig:otherpatterns}(b), Fig.~\ref{fig:impurity_effect}(a) and Fig.~\ref{fig:fragmentation} of the main text, respectively. All other parameters apart from the form of the interaction potential were kept the same as in the listed figures.   
     }
    \label{fig:foster}
\end{figure*}

In Fig.~\ref{fig:foster}, we present the dynamics for three representative quench protocols. Fig.~\ref{fig:foster}(a) replicates the configuration of Fig.~\ref{fig:otherpatterns}(b); Fig.~\ref{fig:foster}(b) follows the impurity protocol of Fig.~\ref{fig:impurity_effect}; and Fig.~\ref{fig:foster}(c) shows the selective quench from Fig.~\ref{fig:fragmentation}. Our results reveal no qualitative differences when transitioning from $r^{-6}$ to $r^{-3}$ interspecies interactions. The spatial structures of the frozen regions, the frequency of the localized oscillations, and the shielding of the central fragments remain qualitatively intact. These results show that the fragmentation phenomenology is robust against a substantial change in the spatial decay of the interspecies interaction, provided the constraint hierarchy is preserved.

\end{document}


\title{Supplementary Material for ``Long-lived revivals and real-space fragmentation in chains of multispecies Rydberg atoms''}

\author{Jose Soto-Garcia}
\affiliation{Rudolf Peierls Centre for Theoretical Physics, University of Oxford, Clarendon Laboratory, Oxford OX1 3PU, United Kingdom}

\author{Natalia Chepiga}
\affiliation{Rudolf Peierls Centre for Theoretical Physics, University of Oxford, Clarendon Laboratory, Oxford OX1 3PU, United Kingdom}

\begin{abstract}
This Supplementary Material provides additional numerical and methodological support for the constrained dynamics reported in the main text. We first connect the dimensionless parameters used in the simulations to experimentally relevant Rb–Cs scales. We then establish the robustness of real-space fragmentation and long-lived revival dynamics against variations in interaction strengths, initial detunings, atomic arrangements, and quench protocols, and show that these features persist to times exceeding the characteristic scale associated with residual longer-range interactions. We further demonstrate species-selective fragmentation in the all-repulsive regime, characterize the double-frequency structure responsible for long-time beating of the revivals, and explore alternative interaction hierarchies, including attractive Cs–Cs interactions supporting both static dynamical barriers and propagating defects. Finally, we quantify the dependence of revival quality and entanglement growth on state preparation and compare the oscillation frequencies of elementary fragments with few-body expectations. These results demonstrate that multispecies Rydberg arrays provide a robust and versatile setting for engineering fragmented many-body dynamics and coherent revivals across experimentally relevant parameter regimes.

\end{abstract}

\maketitle

\section{Experimental parameter scales}

In the main text, we set $\Omega_{\mathrm{Cs}}=\Omega_{\mathrm{Rb}}=1$ and present the dynamics up to $t\Omega=50$. To relate the dimensionless parameters used in our simulations to experimentally relevant scales, we consider representative parameters from
the dual-species Rb--Cs experiment of Anand \textit{et al.}~\cite{anand2024dual}.
For a typical experimental parameter set, the Rydberg levels and Rabi
frequencies are
\begin{align}
\mathrm{Rb}:&\qquad
n=68,
\qquad
\frac{\Omega_{\mathrm{Rb}}}{2\pi}
=2.380(5)\,\mathrm{MHz},
\\
\mathrm{Cs}:&\qquad
n=67,
\qquad
\frac{\Omega_{\mathrm{Cs}}}{2\pi}
=1.860(7)\,\mathrm{MHz}.
\end{align}
The corresponding Rydberg-state lifetimes are approximately
\begin{equation}
\tau_{\mathrm{Rb}}=138\,\mu\mathrm{s},
\qquad
\tau_{\mathrm{Cs}}=126\,\mu\mathrm{s}.
\end{equation}
We emphasize that these Rabi frequencies correspond to a particular
experimental calibration and differ slightly from those employed in the
dual-species blockade measurements shown in Fig.~3 of
Ref.~\cite{anand2024dual}.

Taking a representative common Rabi-frequency scale
\begin{equation}
\frac{\Omega}{2\pi}=2\,\mathrm{MHz},
\end{equation}
gives
\begin{equation}
\Omega^{-1}
=
\frac{1}{2\pi\times2\,\mathrm{MHz}}
\simeq
0.0796\,\mu\mathrm{s}.
\end{equation}
The longest time shown in the main text therefore corresponds to
\begin{equation}
t\Omega=50
\qquad\Longleftrightarrow\qquad
t_{\max}\simeq3.98\,\mu\mathrm{s}.
\end{equation}
For comparison, the oscillation period of a single resonantly driven atom is
\begin{equation}
T_{\mathrm{single}}
=
\frac{2\pi}{\Omega}
\simeq
0.50\,\mu\mathrm{s},
\end{equation}
so the time interval shown in the main-text figures contains approximately
eight single-atom Rabi-oscillation periods.

The simulated time window is also short compared with the bare
Rydberg-state lifetimes:
\begin{equation}
\frac{t_{\max}}{\tau_{\mathrm{Cs}}}
\simeq0.032,
\qquad
\frac{t_{\max}}{\tau_{\mathrm{Rb}}}
\simeq0.029.
\end{equation}
Even under the conservative assumption that an atom remains continuously in
the Rydberg state throughout the complete evolution, its survival probability
over this interval is approximately
\begin{equation}
P_{\mathrm{surv}}(t_{\max})
=
\exp\left(-\frac{t_{\max}}{\tau}\right)
\simeq0.97.
\end{equation}
In the actual many-body dynamics, most sites have a time-dependent Rydberg
population $\langle n_i(t)\rangle<1$, so their integrated Rydberg occupation,
and hence their spontaneous-decay probability, is smaller than this estimate. If one additionally assumes independent decay and
continuous Rydberg occupation for every atom, the corresponding no-decay
probability for a chain of $N$ atoms would be
\begin{equation}
P_{\mathrm{chain}}
=
\left(P_{\mathrm{surv}}\right)^N
\simeq0.97^N.
\end{equation}
For $N=20$, this gives $P_{\mathrm{chain}}\simeq0.54$. This estimate is again
strongly conservative because the atoms are not continuously populated in
the Rydberg state during the evolution.

We next compare the interaction scale used in the simulations with
experimentally accessible interatomic separations. The representative
nearest-neighbour interaction strength employed in the main text is
\begin{equation}
C_0=|C_1|
=
1.2^6\Omega
\simeq
2.99\,\Omega.
\end{equation}
For $\Omega/(2\pi)=2\,\mathrm{MHz}$, this corresponds to
\begin{equation}
\frac{V_{\mathrm{nn}}}{2\pi}
=
1.2^6\frac{\Omega}{2\pi}
\simeq
5.97\,\mathrm{MHz}.
\end{equation}
The dual-species experiment of Ref.~\cite{anand2024dual} demonstrated an
effective Rb--Cs interaction of approximately $24\,\mathrm{MHz}$ at a
separation $r_0=5.6\,\mu\mathrm{m}$. Near the F\"orster resonance, where the
interaction has an approximately dipolar dependence
$V(r)\propto r^{-3}$, an interaction of approximately
$6\,\mathrm{MHz}$ would therefore be obtained at
\begin{align}
a
&=
r_0
\left(
\frac{V(r_0)}{V(a)}
\right)^{1/3}
\nonumber\\
&\simeq
5.6\,\mu\mathrm{m}
\left(
\frac{24\,\mathrm{MHz}}
     {6\,\mathrm{MHz}}
\right)^{1/3}
\simeq
8.9\,\mu\mathrm{m}.
\end{align}
For comparison, if one instead assumes the van der Waals scaling
$V(r)\propto r^{-6}$ employed in our model, the same change in interaction
strength gives
\begin{align}
a
&=
r_0
\left(
\frac{V(r_0)}{V(a)}
\right)^{1/6}
\nonumber\\
&\simeq
5.6\,\mu\mathrm{m}
\left(
\frac{24\,\mathrm{MHz}}
     {6\,\mathrm{MHz}}
\right)^{1/6}
\simeq
7.1\,\mu\mathrm{m}.
\end{align}
Thus, both estimates correspond to interatomic separations accessible in
programmable optical-tweezer arrays.

We stress, however, that the experimentally demonstrated interaction
parameters do not necessarily satisfy the symmetric relation
\begin{equation}
C_{\mathrm{RbRb}}
=
C_{\mathrm{CsCs}}
=
-C_{\mathrm{RbCs}}
\end{equation}
at a single lattice spacing. Our parameter choice should therefore be
understood as an idealized interaction hierarchy rather than as an exact
representation of the specific
$68S_{\mathrm{Rb}}$--$67S_{\mathrm{Cs}}$ experimental configuration.
Nevertheless, the relevant absolute interaction scale,
$V_{\mathrm{nn}}/\Omega\simeq3$, and the associated dynamical timescales
are experimentally accessible.

Finally, the experimentally measured damping times of continuously driven
ground--Rydberg Rabi oscillations are
\begin{equation}
T_{\mathrm{damp,Rb}}
=
4.1(4)\,\mu\mathrm{s},
\qquad
T_{\mathrm{damp,Cs}}
=
4.6(6)\,\mu\mathrm{s}.
\end{equation}
These damping times are substantially shorter than the bare Rydberg-state
lifetimes and are comparable to the longest simulated time,
$t_{\max}\simeq4\,\mu\mathrm{s}$. This comparison indicates that spontaneous
Rydberg decay should remain weak over the simulated interval, whereas
technical dephasing associated with laser noise, atomic motion, and spatial
inhomogeneities may reduce the contrast of the oscillations at the latest
times. Nevertheless, the experimentally measured coherence times are
comparable to the full main-text simulation window and should allow the
formation of the frozen regions and several revival cycles to be resolved.

\section{Robustness and long-time dynamics}
\label{sec:robustness}

In the main text we choose a representative interaction
strength, typically taking the prefactors of the van der Waals
interactions, $C_{\alpha\beta}$, with \(\alpha,\beta\in\{\mathrm{Cs},\mathrm{Rb}\}\), to be of order $1.2^6\,\Omega$. For the interaction hierarchy used in the main text,  $C_\mathrm{CsCs} = C_\mathrm{RbRb} \equiv C_0$ and $C_\mathrm{CsRb} = C_1$. 
This choice facilitates comparison between the different dynamical
protocols at similar interaction scales. The constrained dynamics
reported in the main text, does not rely on this particular parameter choice.
To demonstrate this, here we examine its robustness by
varying both the interaction strengths and the initial detunings, and
by extending representative simulations to substantially longer
times. The parameters used in all supplementary simulations are
summarized in Table~\ref{tab:parameter_values}.

The longer time window is particularly relevant because the van der
Waals interaction is not truncated to nearest neighbours. For the
representative interaction prefactor
\begin{equation}
    C = 1.2^6\Omega,
\end{equation}
the interaction between atoms separated by two lattice spacings is
\begin{equation}
    V_{\mathrm{NNN}}
    =
    \frac{C}{2^6}
    =
    \frac{1.2^6}{2^6}\Omega
    \simeq
    0.0467\,\Omega.
\end{equation}
The corresponding time required for this interaction to accumulate a
phase of order unity is
\begin{equation}
    t_{\mathrm{NNN}}
    \sim
    \frac{1}{V_{\mathrm{NNN}}}
    \simeq
    21.4\,\Omega^{-1},
\end{equation}
while a complete characteristic cycle associated with this energy
scale occurs on the timescale
\begin{equation}
    T_{\mathrm{NNN}}
    =
    \frac{2\pi}{V_{\mathrm{NNN}}}
    \simeq
    135\,\Omega^{-1}.
\end{equation}
The longest simulations presented in the next subsections extend to approximately
$150\,\Omega^{-1}$ and therefore probe times longer than both of these
characteristic scales. They provide a direct test of whether residual
longer-range interactions rapidly destroy the fragmented dynamics
observed at shorter times.

\subsection{Attractive Cs--Rb interactions}
\label{sec:attractive_csrb}

\begin{figure}[ht]
    \centering
    \includegraphics[width=0.82\linewidth]
    {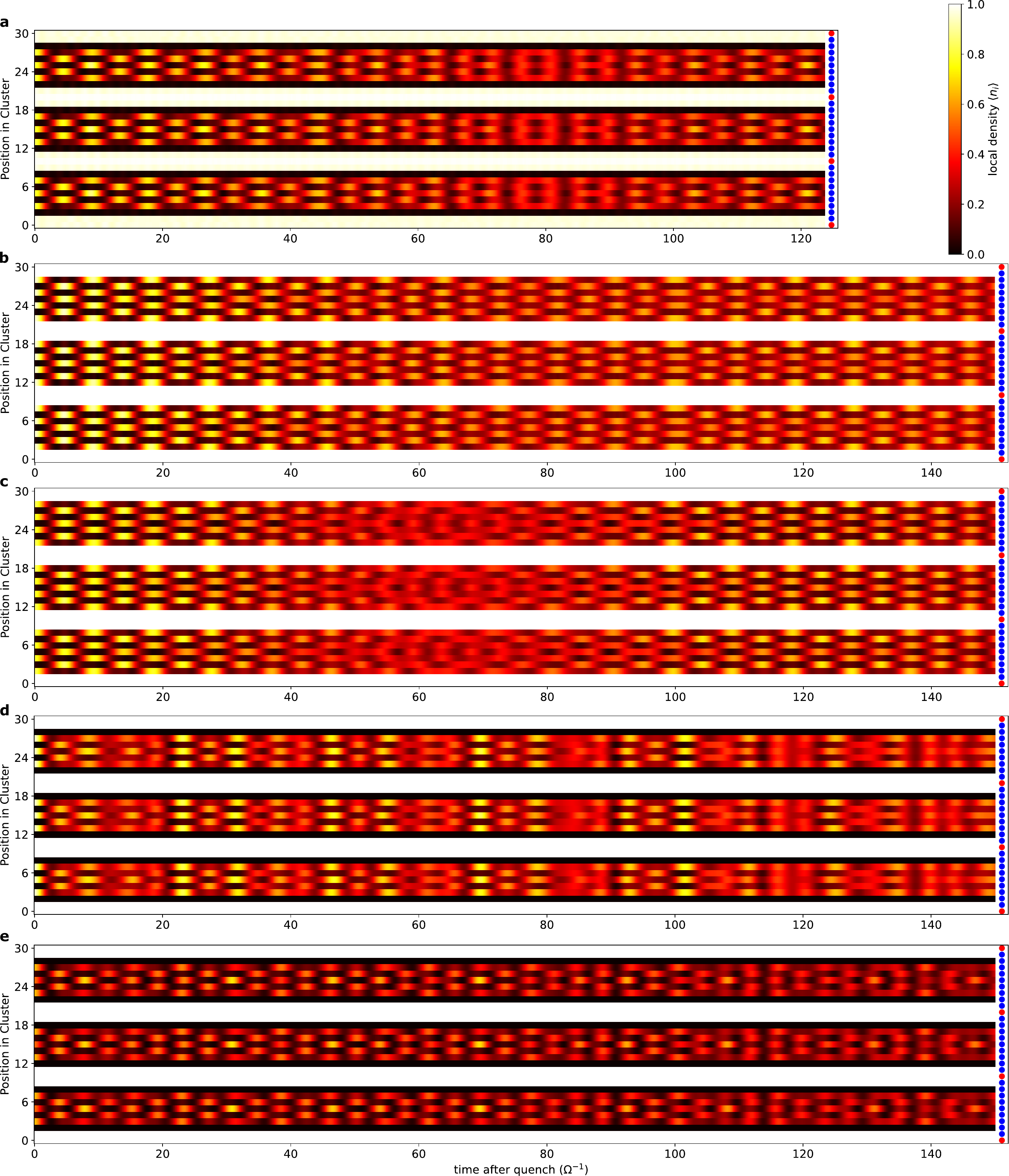}
    \caption{
    Robustness and long-time evolution of the fragmented dynamics for
    attractive Cs--Rb interactions. The colour scale denotes the local
    Rydberg excitation density $\langle n_i(t)\rangle$, while the
    sidebars indicate the atomic arrangement, with red (blue) circles
    denoting Cs (Rb) atoms. Panels (a)--(e) correspond to the parameter
    sets listed in Table~\ref{tab:parameter_values}. In all cases,
    nearly frozen Rb--Cs--Rb clusters separate dynamically active Rb
    regions. In panels (b) and (c), the Rb atoms immediately adjacent
    to these frozen clusters participate in the period-two oscillatory
    dynamics because they are initially populated in the Rydberg state,
    whereas in the remaining panels these sites remain predominantly
    frozen in the ground state. Variations of the microscopic
    interaction strengths modify the oscillation frequencies and the
    detailed revival structure without destroying the fragmented
    spatial pattern. The simulations extend to times comparable to or
    longer than the characteristic period associated with the
    next-nearest-neighbour interaction.
    }
    \label{fig:attractive_interactions}
\end{figure}

The fragmented dynamics discussed in the main text for repulsive Cs--Cs and Rb--Rb interactions and attractive Cs--Rb interactions is robust against substantial changes in the microscopic parameters. Fig.~\ref{fig:attractive_interactions} illustrates this by varying both the interaction strengths and their relative ratios, as well as the initial detunings, while retaining the same qualitative interaction hierarchy. Across all of these representative variations, nearly frozen Rb--Cs--Rb clusters continue to act as dynamical barriers separating Rb regions with coherent oscillatory dynamics. The principal effect of changing the parameters is quantitative: the characteristic oscillation frequencies, revival amplitudes, and detailed interference patterns are modified, whereas the real-space fragmentation itself remains intact.

A notable difference appears in
Figs.~\ref{fig:attractive_interactions}(b) and (c) compared with the
remaining panels of Fig.~\ref{fig:attractive_interactions}, and with
Fig.~\ref{fig:fragmentedp2}(a) of the main text. In the latter cases,
the Rb atoms immediately adjacent to each frozen Rb--Cs--Rb cluster
remain approximately frozen in the ground state and therefore delimit
the dynamically active region. By contrast, in
Figs.~\ref{fig:attractive_interactions}(b) and (c), these neighbouring
Rb atoms participate in the period-two oscillatory dynamics rather
than remaining frozen.

This difference originates from the initial-state resulting from the initial set of parameters. For
the parameter sets corresponding to panels (b) and (c), these Rb sites
are initially populated predominantly in the Rydberg state rather than
in the ground state. Consequently, after the quench they become part
of the oscillating period-two sector instead of remaining frozen. The
dynamically active region is therefore enlarged relative to the
examples in which the Rb atoms neighbouring the frozen Rb--Cs--Rb
cluster remain in the ground state.

Importantly, the simulations in
Fig.~\ref{fig:attractive_interactions} extend well beyond the time
window considered in the corresponding figures of the main text.
The fragmented structure and the associated local oscillations persist
up to $t\simeq150\,\Omega^{-1}$, i.e., beyond the characteristic
timescale $T_{\mathrm{NNN}}\simeq135\,\Omega^{-1}$ associated with the
next-nearest-neighbour interaction for the representative coupling
$C=1.2^6\Omega$. Residual long-range interactions do not rapidly destroy
the separation between frozen and dynamically active regions over the
accessible simulation times. We therefore find no indication of
thermalization on this timescale, although the finite simulation
window does not allow us to make a statement about the asymptotic
$t\rightarrow\infty$ dynamics.

\subsection{Selective quenches with repulsive interactions}
\label{sec:repulsive}

\begin{figure}[ht]
    \centering
    \includegraphics[width=0.70\linewidth]
    {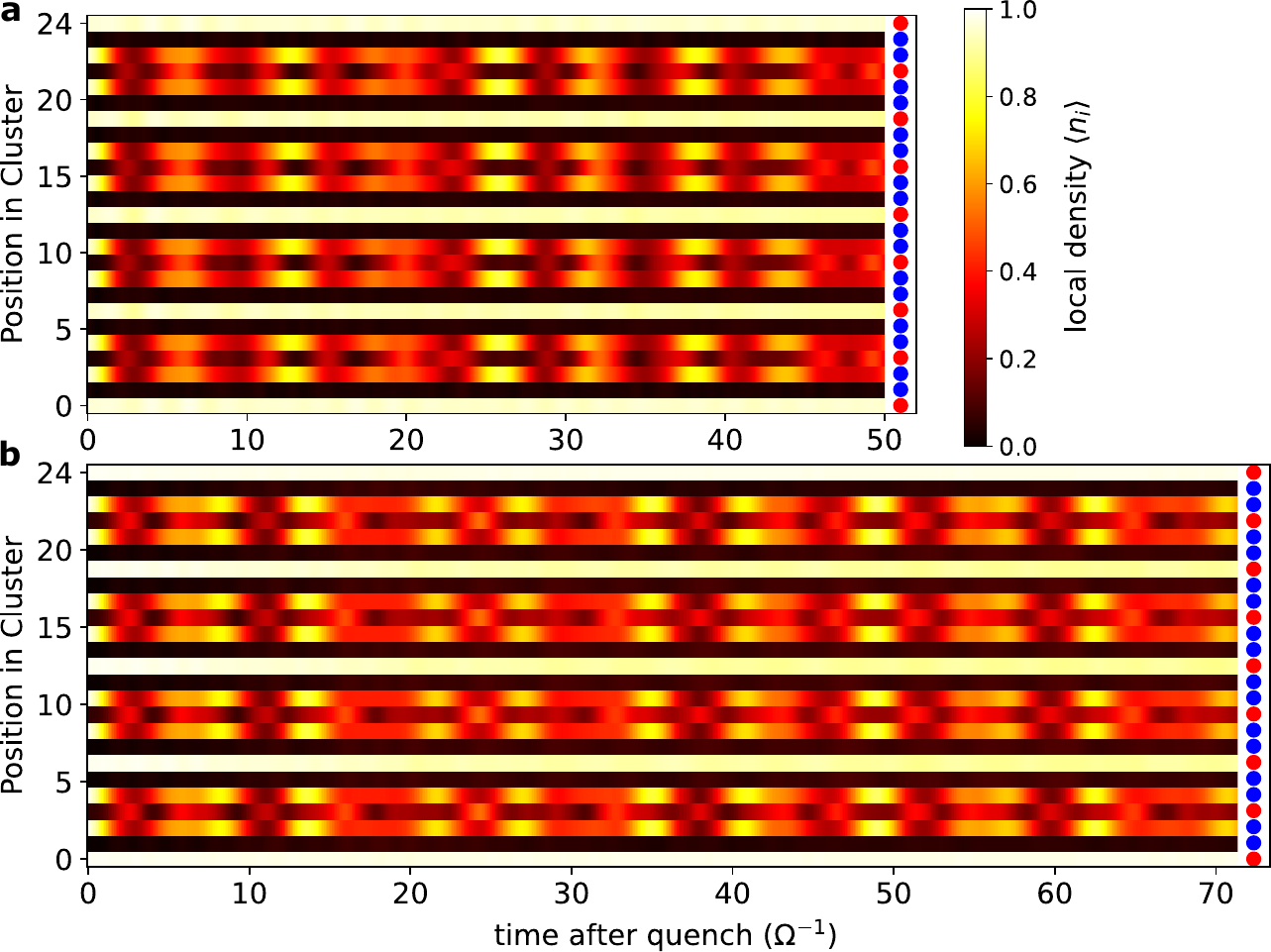}
    \caption{
Robustness of the emergent fragmentation generated by a species-selective quench in the all-repulsive interaction regime. The colour scale shows the local Rydberg excitation density $(\langle n_i(t)\rangle)$, while the sidebars indicate the atomic arrangement, with red (blue) circles denoting Cs (Rb) atoms. Panels (a) and (b) show representative parameter variations relative to the regime considered in the main text; the corresponding values are listed in Table~\ref{tab:parameter_values}. In both cases the selective quench produces dynamically active fragments separated by sites with strongly suppressed dynamics. Changes in the microscopic interaction strengths modify the revival frequencies and detailed temporal structure while preserving the qualitative real-space fragmentation.
    }
    \label{fig:repulsive_interactions}
\end{figure}

The species-selective fragmentation mechanism identified in the main text is likewise not restricted to the particular interaction strengths chosen there. In the all-repulsive regime, keeping one species off resonance while quenching the other dynamically generates sites with strongly suppressed dynamics that separate resonant regions into effective fragments. Fig.~\ref{fig:repulsive_interactions} demonstrates the robustness of this mechanism for representative interaction ratios and initial detunings different from those used in Fig.~4 of the main text. The initial states are crystalline states prepared using the detunings listed in Table~\ref{tab:parameter_values}; at the quench, the Rb detuning is set to zero while the Cs detuning remains unchanged.

For both parameter variations, the evolution develops the same qualitative spatial organization: nearly frozen regions separate dynamically active fragments, within which the local excitation density displays irregular revivals. Changing the microscopic interaction scales modifies the revival frequencies and detailed interference pattern, but not the emergence of the fragmented structure. These examples therefore show that species-selective fragmentation in the all-repulsive regime is robust to appreciable changes in the interaction parameters and is not a fine-tuned feature of the parameter set adopted in the main text.

\subsection{Effects of the initial detunings $\Delta_{\mathrm{Cs},0}$ and $\Delta_{\mathrm{Rb},0}$ on the dynamics}

\begin{figure}[ht]
    \centering
    \includegraphics[width=0.9\linewidth]{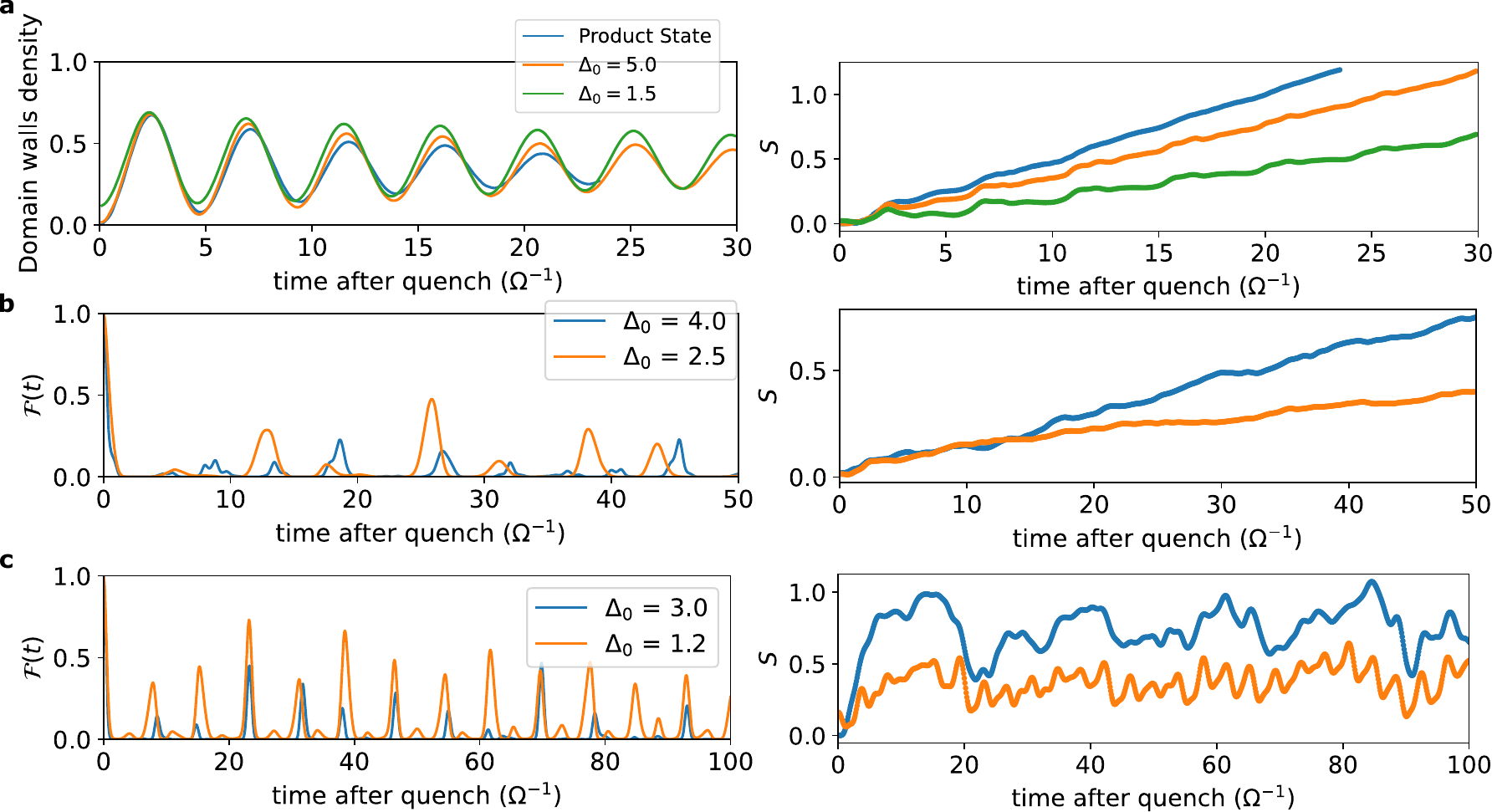}
    \caption{
Dependence of the revival dynamics on the initial detuning
$\Delta_{\mathrm{Cs},0}=\Delta_{\mathrm{Rb},0}\equiv\Delta_0$ for three
representative quench protocols. (a) Single-species Cs chain. The left panel
shows the domain-wall density and the right panel the bipartite entanglement
entropy $S$. (b) Species-selective quench in the all-repulsive interaction
regime. (c) Fragmented dynamics with attractive Cs--Rb interactions. In
(b) and (c), the left panels show the fidelity $\mathcal{F}(t)$ and the right
panels the entanglement entropy. Where applicable,
$|C_0|=|C_1|=1.2^6\Omega$. Smaller initial detunings result in more robust
revivals and slower entanglement growth for the parameter sets considered.
}
\label{fig:delta_effect}
\end{figure}

The initial detunings used in the main text should likewise be regarded as representative choices rather than uniquely selected values required for the appearance of revivals. Varying $\Delta_{\mathrm{Cs},0}$ and $\Delta_{\mathrm{Rb},0}$ changes the state prepared before the quench and therefore modifies the quantitative revival dynamics, but the comparison in Fig.~\ref{fig:delta_effect} shows that the relevant phenomenology persists over a finite range of initial conditions.

Our choice of detunings is partly motivated by the results of Daniel \textit{et al.}~\cite{daniel2023bridging}, who showed, for the PXP model in the presence of a longitudinal field, that quantum many-body revivals can be enhanced for initial states prepared close to an equilibrium quantum critical point. In the multispecies geometries considered here, however, establishing an analogous connection to criticality is considerably less straightforward. In particular, we have not determined the location of a putative critical point for each interaction pattern and atomic arrangement. Moreover, several geometries explicitly break translation symmetry, so that a direct characterization in terms of the phases of a translation-invariant model is not generally applicable. We therefore do not associate the chosen detunings with a well-defined distance from a critical point. Instead, we choose our values of $\Delta_0$ to generate the intended pattern of Rydberg excitations while keeping far from the classical state.

 Fig.~\ref{fig:delta_effect} directly tests the sensitivity of the revival dynamics to the initial detuning for three representative protocols. Across the parameter variations shown, the revivals remain present, while their lifetime and the associated entanglement growth depend quantitatively on $\Delta_0$. Smaller values of $\Delta_0$ lead to more pronounced and longer-lived revivals and slower entanglement growth for the cases considered. Thus, the detunings adopted in the main text are convenient choices within a broader regime supporting revival dynamics, rather than fine-tuned values required for its existence.

\section{Double-frequency structure and beating of the revivals}
\label{sec:double_frequency}

\begin{figure}[t]
    \centering
    \includegraphics[width=0.52\linewidth]
    {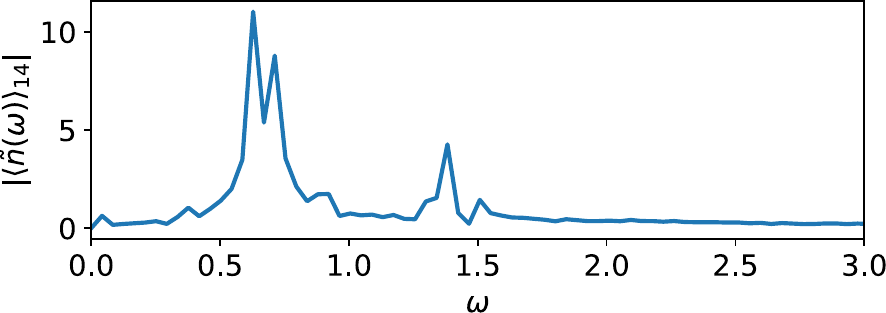}
    \caption{
    Frequency spectrum of the single-site excitation-density dynamics
    corresponding to Fig.~\ref{fig:attractive_interactions}(c).
    The dominant low-frequency contribution is split into two nearby
    components at approximately $\omega_1\simeq0.63\Omega$ and
    $\omega_2\simeq0.71\Omega$. Their mean frequency determines the
    fast oscillatory timescale, whereas their separation produces the
    slower modulation visible in the time-domain dynamics. 
    }
    \label{fig:fourier}
\end{figure}

One of the long-time simulations,
Fig.~\ref{fig:attractive_interactions}(c), exhibits two clearly
separated dynamical timescales. The local excitation density displays
relatively fast revivals whose amplitude is modulated on a much slower
timescale. A similar effect can be observed in Fig.~\ref{fig:impurity_effect}(b) of the main text. To investigate the origin of this
slow modulation, we analyse the Fourier spectrum of the local
excitation density. 
Fig.~\ref{fig:fourier} shows the Fourier transform of the
single-site signal corresponding to
Fig.~\ref{fig:attractive_interactions}(c). Rather than a single
isolated frequency, the low-frequency part of the spectrum contains
two nearby dominant components. The presence of these two frequencies
naturally gives rise to beating in the time domain.

A useful idealized description of quantum many-body scar dynamics is
provided by a set of scarred eigenstates forming an approximately
equally spaced tower,
\begin{equation}
    E_m = E_0 + m\Delta ,
\end{equation}
where $\Delta$ denotes the characteristic energy separation between
consecutive scarred eigenstates. Perfect equal spacing produces
periodic revivals with characteristic period
\begin{equation}
    T_{\mathrm{scar}}
    \simeq
    \frac{2\pi}{\Delta}.
\end{equation}

In a finite interacting many-body system, however, the relevant
spectrum need not be perfectly harmonic. We may schematically write
\begin{equation}
    E_m = E_0 + m\Delta + \delta_m ,
\end{equation}
where $\delta_m$ represents deviations from perfect equal spacing.
The associated transition frequencies are therefore slightly
different. During the time evolution, the corresponding components
accumulate relative phases, leading first to dephasing and later to
partial rephasing. In the time domain this appears as a slow
modulation of the amplitude of the faster revivals.

For the signal in Fig.~\ref{fig:fourier}, the two dominant nearby
low-frequency peaks are located at
\begin{equation}
    \omega_1 = (0.63 \pm 0.04)\,\Omega,
    \qquad
    \omega_2 = (0.71 \pm 0.04)\,\Omega.
\end{equation}
Here the quoted uncertainty is set by the frequency resolution of the
finite-time Fourier transform, as discussed below.

If these two components dominate the oscillatory signal, its leading
time dependence can be represented schematically as
\begin{equation}
    n(t)
    \sim
    \cos(\omega_1 t)+\cos(\omega_2 t).
\end{equation}
Using the trigonometric identity for the sum of two cosines gives
\begin{equation}
    n(t)
    \sim
    2
    \cos\left(
        \frac{\omega_1-\omega_2}{2}t
    \right)
    \cos\left(
        \frac{\omega_1+\omega_2}{2}t
    \right).
\end{equation}
This expression explicitly separates a fast oscillatory component
from a slowly varying envelope.

The fast oscillation is governed by the mean angular frequency
\begin{equation}
    \bar{\omega}
    =
    \frac{\omega_1+\omega_2}{2},
\end{equation}
which gives
\begin{equation}
    T_{\mathrm{fast}}
    \simeq
    \frac{2\pi}{\bar{\omega}}
    =
    \frac{4\pi}{\omega_1+\omega_2}
    =
    (9.4\pm0.4)\,\Omega^{-1}.
\end{equation}
This estimate agrees with the period obtained directly from the
time-domain signal by averaging the separation between successive
maxima,
\begin{equation}
    T_{\mathrm{fast}}^{(\mathrm{time})}
    \simeq
    (9.1\pm0.3)\,\Omega^{-1}.
\end{equation}
The agreement supports the identification of the two Fourier
components with the oscillatory dynamics observed in
Fig.~\ref{fig:attractive_interactions}(c).

The slower amplitude modulation is controlled by the frequency
difference
\begin{equation}
    \delta\omega
    =
    |\omega_1-\omega_2|.
\end{equation}
The magnitude of the beating envelope therefore repeats on the
timescale
\begin{equation}
    T_{\mathrm{env}}
    \simeq
    \frac{2\pi}{|\omega_1-\omega_2|}
    \simeq
    79\,\Omega^{-1}.
\end{equation}
This value is of the same order as the slow modulation visible in the
long-time density evolution.

Because the separation between the two peaks is only a few Fourier
frequency bins, its uncertainty is comparatively large. We therefore
do not regard the numerical value of $T_{\mathrm{env}}$ as a precise
measurement. Rather, the Fourier analysis establishes that the
long-time signal contains nearby frequencies whose interference is
consistent with the observed beating. Such a splitting is also
consistent with deviations from a perfectly equally spaced spectrum
of the states contributing to the revivals.

\subsection{Frequency resolution}
\label{sec:fourier_resolution}

For a discrete time series sampled at intervals $\delta t$ and containing
$N$ points, the total observation time is
\begin{equation}
    T_{\mathrm{obs}} = N\,\delta t ,
\end{equation}
and the spacing between consecutive angular-frequency bins of the
discrete Fourier transform is
\begin{equation}
    \Delta\omega_{\mathrm{FFT}}
    =
    \frac{2\pi}{N\,\delta t}
    =
    \frac{2\pi}{T_{\mathrm{obs}}}.
\end{equation}

For the time window used in Fig.~\ref{fig:fourier}, this gives
\begin{equation}
    \Delta\omega_{\mathrm{FFT}}
    \simeq
    0.04\,\Omega.
\end{equation}
We use this frequency-bin spacing as an estimate of the resolution
with which the positions of the Fourier peaks can be identified. It
should not be interpreted as a statistical error bar. This distinction
is particularly relevant here because the separation
$|\omega_1-\omega_2|$ is only a few times the Fourier-bin spacing.

\section{Attractive intraspecies Cs--Cs interactions}
\label{sec:new_interactions}

The constrained dynamics discussed in the main text is also not restricted to the particular sign structure of the interactions chosen there. Multispecies arrays allow the relative signs and strengths of the intra- and interspecies couplings to be modified, providing a further test of how general the fragmentation mechanism is. As a representative example, we consider an interaction hierarchy in which the Cs--Cs interaction is attractive,
\begin{equation}
    C_{\mathrm{CsCs}}<0,
\end{equation}
while the Rb--Rb and Cs--Rb interactions remain repulsive.

\begin{figure}[ht]
    \centering
    \includegraphics[width=0.95\linewidth]
    {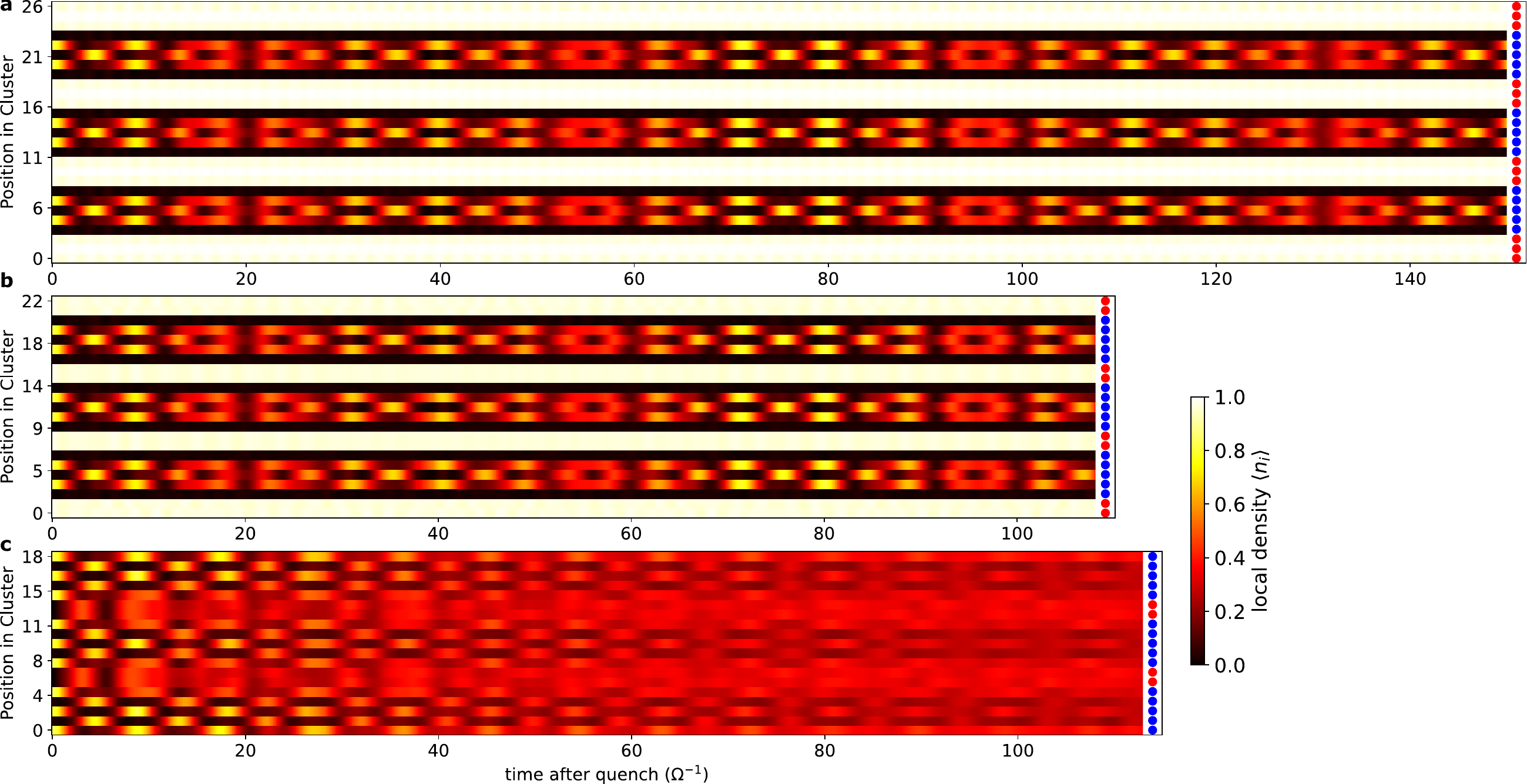}
    \caption{
    Constrained dynamics with attractive Cs--Cs interactions and
    repulsive Rb--Rb and Cs--Rb interactions. (a) Attractive Cs--Cs
    interactions stabilize Cs triplets that act as dynamical barriers
    separating oscillatory Rb regions. (b) The same qualitative
    mechanism persists when the barriers are formed by Cs pairs.
    (c) For a large negative initial Cs detuning, the Cs atoms are
    initially predominantly in the ground state; after the quench, the
    corresponding perturbation becomes mobile and progressively
    affects the surrounding Rb dynamics. These examples demonstrate
    that both static fragmentation and propagating-impurity dynamics
    can be obtained with an interaction hierarchy different from that
    considered in the main text. Parameters for all panels are given
    in Table~\ref{tab:parameter_values}.
    }
    \label{fig:new_interactions}
\end{figure}

Fig.~\ref{fig:new_interactions} shows that qualitatively similar constrained dynamics to the one observed in Fig.~\ref{fig:fragmentedp2} of the main text survives this change of interaction hierarchy. In particular, the species responsible for forming the dynamical barriers can be interchanged, while the basic distinction between localized structures and dynamically active regions remains.

In Fig.~\ref{fig:new_interactions}(a), groups of three neighbouring
Cs atoms form stable high-density structures separating extended Rb
regions. The Cs triplets consequently act as dynamical barriers, and
the intervening Rb sectors exhibit persistent oscillatory dynamics.
Fig.~\ref{fig:new_interactions}(b) shows the analogous situation for
Cs pairs. Although the barrier is formed by a smaller number of Cs
atoms, the same qualitative decomposition into weakly evolving Cs
structures and oscillatory Rb regions remains apparent.

The initial condition is qualitatively different in
Fig.~\ref{fig:new_interactions}(c). Here a large negative initial Cs
detuning, $\Delta_{\mathrm{Cs}}=-10\Omega$, suppresses the Cs
excitation in the initial state. After the quench, the Cs sector no
longer forms a static high-density barrier. Instead, its dynamics
generates a mobile perturbation of the Rb background, producing a
spreading spatiotemporal pattern. This behaviour is analogous to the
propagating-impurity dynamics discussed in the main text for an
initial period-two Rb state. Thus, whether a multispecies structure
acts as a static dynamical barrier or as a mobile defect depends not
only on the interaction hierarchy, but also crucially on the initial
state from which the quench is performed.

\section{Additional dynamical regimes}
\label{sec:new_dynamics}

\begin{figure}[t]
    \centering
    \includegraphics[width=0.4\linewidth]
    {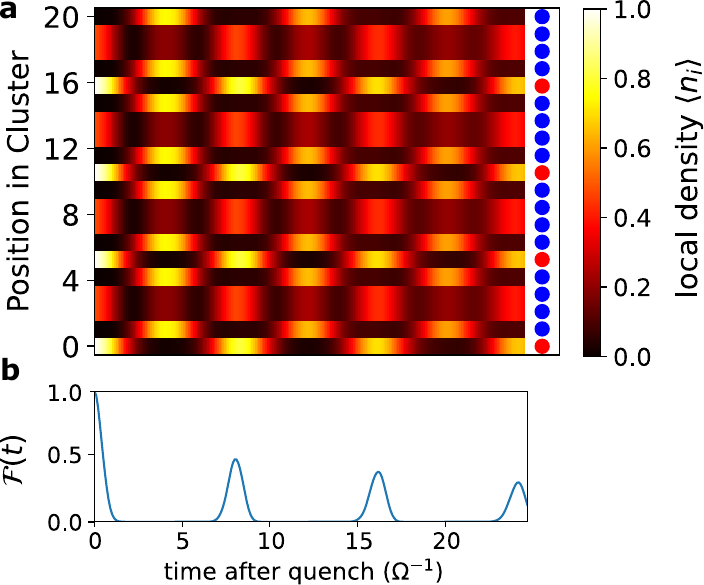}
    \caption{
    Example of constrained many-body dynamics beyond the elementary
    few-body fragments discussed in the main text.
    (a) Time evolution of the local Rydberg excitation density in a
    Cs--Rb$_3$ array following a global quench from
    $\Delta_{\mathrm{Cs},0}=5\Omega$ and
    $\Delta_{\mathrm{Rb},0}=2.5\Omega$ to
    $\Delta_{\mathrm{Cs}}=\Delta_{\mathrm{Rb}}=0$, with
    $C_{\mathrm{CsCs}}=C_{\mathrm{RbRb}}=C_{\mathrm{CsRb}}
    =1.4^6\Omega$.
    Different sites exhibit distinct oscillation amplitudes and phases,
    resulting in a spatially structured density pattern.
    (b) Fidelity $\mathcal{F}(t)$ for the same evolution, displaying
    recurrent many-body revivals. Periodic boundary conditions are imposed to suppress boundary-induced effects that would otherwise generate propagating disturbances and degrade the revival structure.
    }
    \label{fig:new_dynamics}
\end{figure}

The combination of species-dependent interactions, programmable
atomic arrangements, and independent control of the two laser
detunings enables constrained dynamics beyond the elementary
few-body fragments discussed in the main text. Fig.~\ref{fig:new_dynamics}
shows one such example in a Cs--Rb$_4$ array with all-repulsive
interactions,
\[
C_{\mathrm{CsCs}}
=
C_{\mathrm{RbRb}}
=
C_{\mathrm{CsRb}}
=
1.4^6\Omega .
\]
The initial state is prepared using different detunings for the two
species, $\Delta_{\mathrm{Cs},0}=5\Omega$ and
$\Delta_{\mathrm{Rb},0}=2.5\Omega$. 
This choice produces a structured
initial configuration in which highly excited Cs sites coexist with
resonant Rb excitation pairs. At $t=0$, both detunings are quenched to
resonance, $\Delta_{\mathrm{Cs}}=\Delta_{\mathrm{Rb}}=0$.
For this example, we impose periodic boundary conditions in order to suppress boundary-induced effects. With open boundaries, the edge sites exhibit different oscillation periods, which generate propagating disturbances that progressively disrupt the revival structure. As shown in Fig.~\ref{fig:new_dynamics}(a), the subsequent evolution comprises different sites exhibiting distinct oscillation
amplitudes, producing a spatially modulated excitation
pattern across the array. This behavior reflects the simultaneous
evolution of several locally constrained degrees of freedom generated
by the multispecies geometry and by the correlations already present
in the initial state.

Despite this nonuniform local dynamics, the many-body evolution
retains a pronounced collective structure. The fidelity shown in
Fig.~\ref{fig:new_dynamics}(b) exhibits recurrent revivals at
approximately regular intervals, demonstrating repeated partial
returns of the system towards its initial state. 

\section{Frequency oscillations: single atoms and pairs}

The elementary dynamically active sectors identified in Fig.~1 of the main text provide a useful reference for the characteristic frequencies appearing throughout the fragmented dynamics. For the Cs--Rb$_5$ configuration, the frozen Rb--Cs--Rb regions isolate a single active Rb atom, while for Cs--Rb$_6$ they isolate a pair of neighbouring Rb atoms sharing a resonant single Rydberg excitation. In the ideal isolated limit, the corresponding oscillation frequencies are
\begin{equation}
    \omega_{\mathrm{single}}=\Omega,
    \qquad
    \omega_{\mathrm{pair}}=\sqrt{2}\,\Omega,
\end{equation}
or equivalently
\begin{equation}
    T_{\mathrm{single}}=\frac{2\pi}{\Omega},
    \qquad
    T_{\mathrm{pair}}=\frac{2\pi}{\sqrt{2}\Omega}.
\end{equation}
As discussed in the main text, the numerically extracted periods for the representative parameters are
$T_{\mathrm{single}}\simeq 6.37\pm0.05\,\Omega^{-1}$ and
$T_{\mathrm{pair}}\simeq 4.42\pm0.15\,\Omega^{-1}$, in close agreement with these ideal few-body predictions. In particular, the pair dynamics displays the expected $\sqrt{2}$ enhancement of the effective Rabi frequency.

\begin{figure}[ht]
    \centering
    \includegraphics[width=0.8\linewidth]{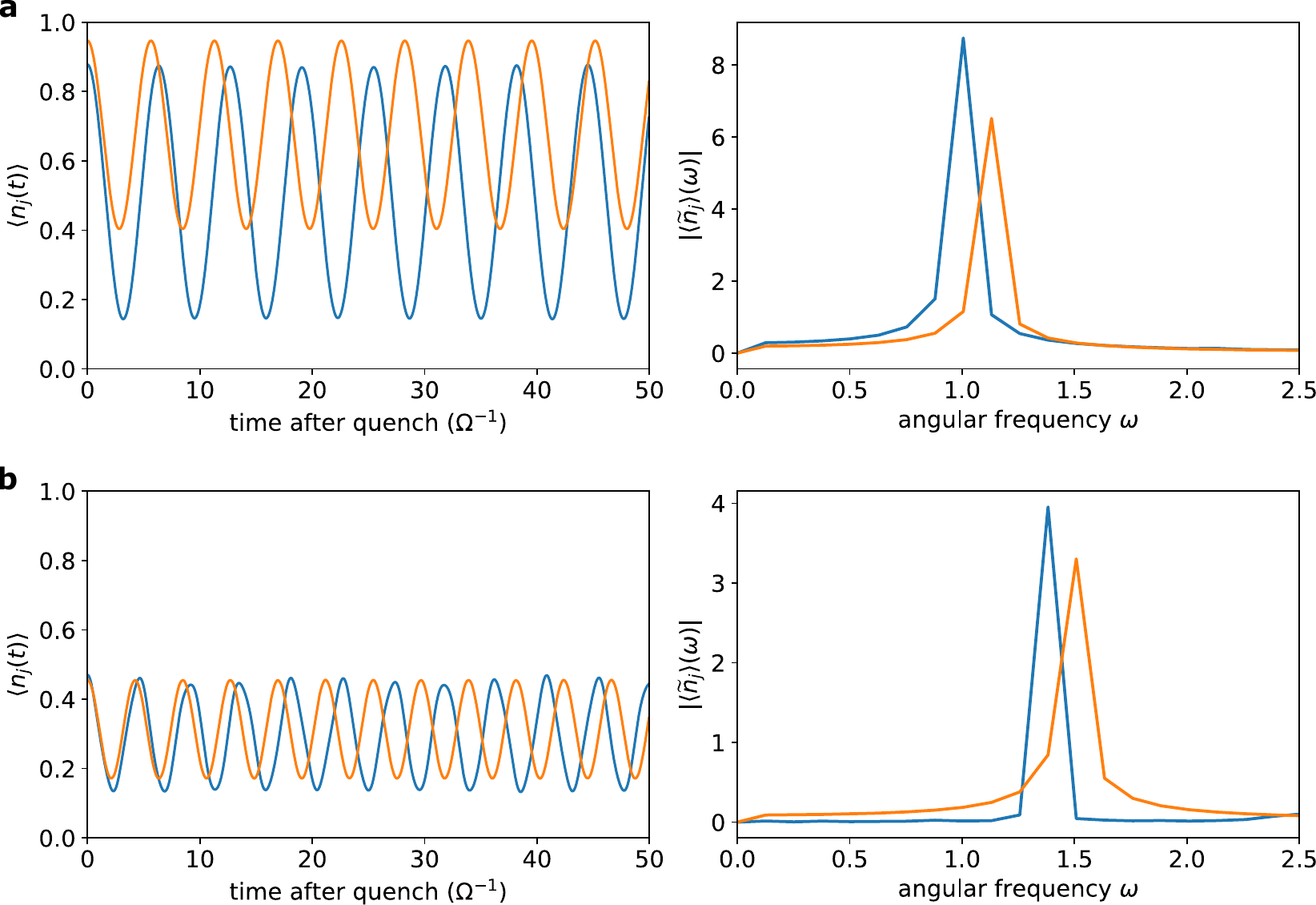}
    \caption{Local excitation dynamics and corresponding Fourier spectra for the elementary dynamically active fragments identified in Fig.~1 of the main text. (a) Single active Rb atom and (b) resonant pair of neighbouring Rb atoms. Blue curves correspond to
    $\Delta_{\mathrm{Cs},0}=\Delta_{\mathrm{Rb},0}=1.2\,\Omega$ and
    $C_{\mathrm{CsCs}}=C_{\mathrm{RbRb}}=-C_{\mathrm{CsRb}}=1.2^6\,\Omega$,
    while orange curves correspond to
    $\Delta_{\mathrm{Cs},0}=\Delta_{\mathrm{Rb},0}=1.5\,\Omega$ and
    $C_{\mathrm{CsCs}}=C_{\mathrm{RbRb}}=-C_{\mathrm{CsRb}}=1.5^6\,\Omega$.
    In both panels, the parameter set with more strongly suppressed population transfer toward the ground state exhibits a shift of the dominant Fourier peak to higher frequency. The characteristic frequencies remain close to those expected for an isolated single atom, $\omega_{\mathrm{single}}=\Omega$, and a resonant pair, $\omega_{\mathrm{pair}}=\sqrt{2}\Omega$, respectively.}
    \label{fig:single_pair}
\end{figure}

In Fig.~\ref{fig:single_pair}, we further compare these elementary oscillations for two choices of the interaction strength and initial detuning, using the same interaction convention as in the main text. The blue curves correspond to
\begin{equation}
    \Delta_{\mathrm{Cs},0}
    =
    \Delta_{\mathrm{Rb},0}
    =
    1.2\,\Omega,
    \qquad
    C_{\mathrm{CsCs}}
    =
    C_{\mathrm{RbRb}}
    =
    -C_{\mathrm{CsRb}}
    =
    1.2^6\,\Omega,
\end{equation}
whereas the orange curves correspond to
\begin{equation}
    \Delta_{\mathrm{Cs},0}
    =
    \Delta_{\mathrm{Rb},0}
    =
    1.5\,\Omega,
    \qquad
    C_{\mathrm{CsCs}}
    =
    C_{\mathrm{RbRb}}
    =
    -C_{\mathrm{CsRb}}
    =
    1.5^6\,\Omega.
\end{equation}
In both cases, the detunings are quenched to
$\Delta_{\mathrm{Cs}}=\Delta_{\mathrm{Rb}}=0$ at $t=0$, as in the global-quench protocol considered in the main text. The sign structure corresponds to repulsive intraspecies and attractive interspecies interactions.

For both the single active atom and the resonant pair, the oscillations exhibit incomplete population transfer compared with the corresponding ideal isolated few-body systems. For the single atom, the excitation density does not span the full interval from zero to one, while for the resonant pair the excitation density per atom remains below the ideal maximum and does not approach the ground-state limit completely. This incomplete transfer is consistent with the residual influence of the surrounding frozen regions discussed in the main text.

The comparison between the two parameter sets reveals the same trend in both elementary fragments. For the orange curves, the population transfer toward the ground state is more strongly suppressed than for the blue curves. At the same time, the corresponding Fourier spectra show that the dominant oscillation peak is shifted to a higher angular frequency. Thus, for both the single-atom and resonant-pair sectors, a smaller population transfer toward the ground state is accompanied by a higher characteristic oscillation frequency. The simultaneous modification of the oscillation amplitude and frequency indicates that, although the fragments retain the characteristic single-atom and collective-pair dynamics, their effective few-body evolution is renormalized by the surrounding interacting environment.

\section{Parameters used in the supplementary simulations}
\label{sec:parameters}

For completeness, Table~\ref{tab:parameter_values} summarizes the parameters used in the supplementary simulations. We denote by $\Delta_{\alpha,0}$ the detuning used to prepare the initial state and by $\Delta_{\alpha,f}$ the detuning governing the post-quench evolution, with $\alpha\in\{\mathrm{Cs},\mathrm{Rb}\}$. All detunings are given in units of $\Omega$. The interaction parameters $C_{\mathrm{CsCs}}$, $C_{\mathrm{RbRb}}$, and $C_{\mathrm{CsRb}}$ denote the prefactors of the corresponding van der Waals interactions,
\begin{equation}
V_{\alpha\beta}(r)=\frac{C_{\alpha\beta}}{r^6}.
\end{equation}
Positive (negative) values correspond to repulsive (attractive) interactions.

\begin{table}[ht]
\centering
\caption{
Parameters used for the simulations shown in the Supplementary
Material. $\Delta_{\alpha,0}$ denotes the detuning used for
initial-state preparation and $\Delta_{\alpha,f}$ the detuning after
the quench. The last three columns give the interaction prefactors for
the Cs--Cs, Rb--Rb, and Cs--Rb van der Waals interactions,
respectively.
}
\begin{tabular}{cccccccc}
\hline
Figure
& $\Delta_{\mathrm{Cs},0}/\Omega$
& $\Delta_{\mathrm{Rb},0}/\Omega$
& $\Delta_{\mathrm{Cs},f}/\Omega$
& $\Delta_{\mathrm{Rb},f}/\Omega$
& $C_{\mathrm{CsCs}}/\Omega$
& $C_{\mathrm{RbRb}}/\Omega$
& $C_{\mathrm{CsRb}}/\Omega$
\\
\hline

\ref{fig:attractive_interactions}(a)
& 1.2 & 1.2 & 0 & 0
& $1.2^6$ & $1.2^6$ & $-(1.2)^6$
\\

\ref{fig:attractive_interactions}(b)
& 3 & 3 & 0 & 0
& $1.2^6$ & $1.2^6$ & $-(2.4)^6$
\\

\ref{fig:attractive_interactions}(c)
& 1.2 & 1.2 & 0 & 0
& $1.2^6$ & $1.2^6$ & $-(2.4)^6$
\\

\ref{fig:attractive_interactions}(d)
& 3 & 3 & 0 & 0
& $1.8^6$ & $1.8^6$ & $-(2.4)^6$
\\

\ref{fig:attractive_interactions}(e)
& 1.2 & 1.2 & 0 & 0
& $1.8^6$ & $1.8^6$ & $-(2.4)^6$
\\

\ref{fig:repulsive_interactions}(a)
& 2.5 & 2.5 & 2.5 & 0
& $1.2^6$ & $1.2^6$ & $1.2^6$
\\

\ref{fig:repulsive_interactions}(b)
& 4.0 & 4.0 & 4.0 & 0
& $2.0^6$ & $1.6^6$ & $1.2^6$
\\

\ref{fig:delta_effect}(a) (blue)
& - & - & 0 & -
& $1.2^6$ & - & -
\\

\ref{fig:delta_effect}(a) (orange)
& 5 & - & 0 & -
& $1.2^6$ & - & -
\\

\ref{fig:delta_effect}(a) (green)
& 1.2 & - & 0 & -
& $1.2^6$ & - & -
\\

\ref{fig:delta_effect}(b) (blue)
& 4.0 & 4.0 & 4.0 & 0
& $1.2^6$ & $1.2^6$ & $1.2^6$
\\

\ref{fig:delta_effect}(b) (orange)
& 2.5 & 2.5 & 2.5 & 0
& $1.2^6$ & $1.2^6$ & $1.2^6$
\\

\ref{fig:delta_effect}(c) (blue)
& 3.0 & 3.0 & 0 & 0
& $1.2^6$ & $1.2^6$ & $-(1.2)^6$
\\

\ref{fig:delta_effect}(c) (orange)
& 1.2 & 1.2 & 0 & 0
& $1.2^6$ & $1.2^6$ & $-(1.2)^6$
\\

\ref{fig:new_interactions}(a)
& 1.2 & 1.2 & 0 & 0
& $-(1.2)^6$ & $1.2^6$ & $1.2^6$
\\

\ref{fig:new_interactions}(b)
& 1.2 & 1.2 & 0 & 0
& $-(1.2)^6$ & $1.2^6$ & $1.2^6$
\\

\ref{fig:new_interactions}(c)
& -10.0 & 1.2 & 0 & 0
& $-(1.2)^6$ & $1.2^6$ & $1.2^6$
\\

\ref{fig:new_dynamics}(a)
& 5 & 2.5 & 0 & 0
& $1.4^6$ & $1.4^6$ & $1.4^6$
\\

\ref{fig:single_pair}(a) (blue)
& 1.2 & 1.2 & 0 & 0
& $1.2^6$ & $1.2^6$ & $-(1.2)^6$
\\

\ref{fig:single_pair}(a) (orange)
& 1.5 & 1.5 & 0 & 0
& $1.5^6$ & $1.5^6$ & $-(1.5)^6$
\\

\ref{fig:single_pair}(b) (blue)
& 1.2 & 1.2 & 0 & 0
& $1.2^6$ & $1.2^6$ & $-(1.2)^6$
\\

\ref{fig:single_pair}(b) (orange)
& 1.5 & 1.5 & 0 & 0
& $1.5^6$ & $1.5^6$ & $-(1.5)^6$
\\

\hline
\end{tabular}
\label{tab:parameter_values}
\end{table}

\bibliography{bibliography}